\documentclass[<options>,12pt]{elsarticle}%%{revtex4}
\begin{document}
%\tableofcontents
\title{The vibration of  a beam  with a local unilateral elastic contact}
\date{}
\author[unice]{H.Hazim\corref{cor1}}%%\fnref{fn1}}
\ead{hamad.hazim@unice.fr}
\author[isvr]{N.Ferguson\corref{cor2}}%%\fnref{fn1,fn3}}
%\ead{nsf@isvr.soton.ac.uk}
\author[unice]{B.Rousselet}%%\fnref{fn2}
%\ead{br@unice.fr}
\cortext[cor1]{Corresponding author. tel: +33 04 92076276}
%%\cortext[cor2]{Principal corresponding author}
%%\fntext[fn1]{This is the specimen author footnote.}
%%\fntext[fn2]{Another author footnote, but a little more longer.}
%%\fntext[fn3]{Yet another author footnote. Indeed, you can have
%%any number of author footnotes.}
\address[unice]{J.A.D. Laboratory U.M.R. CNRS 6621, University of Nice Sophia-Antipolis, 06108 Valrose, Nice, France.}
\address[isvr]{Institute of Sound and Vibration Research, University Road, Highfield, Southampton S017 1BJ, UK}
%%\address{
%%\maketitle
\begin{abstract}
The mass reduction of satellite solar arrays results in significant panel flexibility. When such structures are launched in a packed configuration there is a possible striking one with another dynamically, leading ultimately to structural damage during the launch stage. To prevent this, rubber snubbers are mounted at well chosen points of the structure and they act as a one sided linear spring. A negative consequence is that the dynamics of these panels becomes nonlinear.
In this paper a solar array and a snubber are simply modelled as a linear Euler-Bernoulli beam with a one sided linear spring respectively.\\
A numerical and an experimental study of a beam striking a one-sided spring under harmonic excitation is presented.
A finite element model representation is used to solve the partial differential equations governing the structural
dynamics. The models are subsequently validated and updated with experiments.
\end{abstract}
\date{}

\maketitle
\newpage
%%%%%%%%%%%%%%%%%%%%%%%%%%%%%%%%%%%%%%%%%%%%%%%%%%%%%%%%%%%%%%%%%%%%%%%
\section{Introduction}
The study of the nonlinear behaviour of structures with a nonlinear contact or support is a relatively new research field of interest for many structural dynamicists. It is a branch of nonlinear dynamics with a special form of nonlinearity; the system has two linear local components and the nonlinearity comes from the interaction between one with the other. It is a non differentiable nonlinearity and it could be non continuous if one or two components has a strong damping coefficient. Some papers have been published to study such systems \cite{r1,r4,r5}, where the focus was to study the stability using sweep tests experimentally and comparing numerical simulations, the latter computation using special packages for nonlinear simulations.\\
In general, non linear dynamics is a very interesting area of modern research for many reasons; the limited application of the linear theory being one of them. The complexity of the systems studied and used in the new generation of space structures and many other mechanical systems, needs a theory which can deal with the nonlinear behaviour encountered.  Unfortunately, there is no complete theory for nonlinear systems such as for the linear case, but there exists many studies which could be applied for many particular cases by themselves  and from a particular point of view. The interest of the authors was to study the nonlinear systems in both the frequency and the time domains, as well as the internal properties of the systems like nonlinear normal modes (NNM) which is an extension of the well known linear normal modes (LNM) (see  \cite{a3,a4,KR1,P1,JB}) .\\
 \\
The objectives of the current study is to simulate the dynamics of a beam  under periodic excitation when it strikes a linear spring. A finite element numerical model was produced and was validated with subsequent experimental tests.
\\
%%Both numerical and experimental approaches are studied and completed. The numerical modeling and the experimental tests have been done in parallel, signal processing have %%been used to extract the information needed without using black box routines, which could lose the nonlinearity being investigated.\\
%% \\
The study of the total  dynamic behaviour of solar arrays in a folded position with snubbers  are so complicated (see Figure 2), that to simplify, a solar array is modeled by a clamped-free Bernoulli beam with a one-sided linear spring. This system is subjected to a periodic excitation force. The real configuration of the problem is similar to a beam with a unilateral contact subjected to a periodic imposed displacement of the base, but the dynamical behaviour of the system does not change  significantly if the imposed displacement is replaced by a periodic force excitation. The configuration used was easiest to be realized from a technical point of view as the experimental validation rig is very simple to build.\\
 \\
The experimental setup and the rig are briefly presented. The numerical results are also presented and studied in both the frequency and the time domains. It is expected that some similarities to a linear system behaviour will be observed.\\
The effect of the spring location has also been studied. It is important to look for particular points to locate the spring, the aim being to reduce the nonlinear effect as much as possible. The spring was introduced at a point corresponding to the node of the second linear beam mode. In this case it is expected that the system will show a linear behaviour for an excitation near the second natural frequency; however a nonlinear behaviour is expected for different excitation frequencies though.
Note that no signal analysis is done by the acquisition system, as the problem is nonlinear and the standard transfer function calculation is only really applicable and useful for linear systems. The time signal was acquired and the processing performed using external software (Scilab \cite{a6}). The numerical predictions are compared to experimental results and show very good agreement.
\vspace{1cm}
% \begin{comment}
% \begin{figure}[hbtp]
% \begin{center}
% \includegraphics[width=6cm,height=6.5cm]{sat.eps}%%[width=4cm]
% \hspace{0.1cm}
% \includegraphics[width=6.5cm,height=6.5cm]{solar_array.eps}
% \label{solar1}
% \end{center}
% \caption{\small{Left: Solar array of a satellite under a test on a shaker. Right: A solar array from the folded to the final position}}
% \end{figure}
% \end{comment}
%%\newpage
\section{Numerical  modelling}\label{sec1}
% \begin{comment}
% \begin{figure}[hbtp]
% \begin{center}
% \includegraphics[width=14cm,height=7cm]{beam_system_2.eps}
% \end{center}
% \caption{\small{beam system with an unilateral spring under a periodic excitation}}
% \label{1}
% \end{figure}
% \end{comment}
The present study simulated the behaviour of a beam which strikes a snubber under a periodic excitation. As the frequency range of  interest was to consider the first three linear eigen frequencies, the beam was modelled using ten linear Euler Bernoulli beam finite elements. The numerical simulations are presented and compared in the frequency domain. The Fast Fourier Transform was applied to the predicted and the experimental displacements at the free end, i.e. corresponding to the last node of the beam finite element model. The mass effect of the force transducer used for experimental validation was also taken into account in the finite element model.\\
The beam equation of motion with an elastic snubber can be expressed as:
\begin{equation}
\rho S\ddot{u}(x,t)+EIu^{(iv)}(x,t)=F(t)\delta_{x_0}-(k_r u(x_1,t)_-)\delta_{x_1}\label{equation1}
\end{equation}
where $\rho$, $S$, $E$, $I$, $F$ and $k_r$ are respectively the beam density, cross-sectional area, Young's modulus of elasticity, second moment of area, point applied harmonic force at position $x_0$ and an elastic spring attached at position $x_1$.\\
Cantilevered beam boundary conditions assume zero displacement and slope at the fixed end and zero bending moment at the free end. When the elastic unilateral spring is in contact then a shear force is present due to the reaction from the spring,
$$u(0,t)=0,\;\;\partial_x u(0,t)=0;\;\;u(x,0)=0,\;\;\partial_t u(x,0)=0,\partial^2_x u(l,t)=0.$$
The compression of the spring is given by
\begin{equation}
u(x,t)_-=\left\lbrace
\begin{array}{rl}
u(x,t) & \ if \ u \leq 0\\
0  & \ if \ u > 0\\
\end{array} \right.
\end{equation}
The classical Hermite cubic finite element approximation was used to solve the PDE (see \cite{G}), it yields an ordinary nonlinear differential system in the form:
\begin{equation}
M\ddot{q}+Kq=-[k_{r}(q_{n_1})_-]e_{n_1}+F(t) e_{n_2}
\end{equation}
where $M$ and $K$ are respectively the assembled mass and stiffness  matrices, $q$ is the vector of degrees of freedom of the beam,
$q_i=(u_i,\partial_xu_i)$, $i=1,...,n$, where $n$ is the size of $M$, $n_1$ and $n_2$ are the indices of the nodes where the spring and the excitation force are applied to the beam respectively. Numerical time integration was performed using an ODE numerical integration for 'stiff' problems, the package ODEPACK was used  based on the BDF method (backward differentiation formula, see \cite{a6}).
Small damping was  introduced in the spring, there is no damping assumed in the beam structure.\\
\section{Experimental validation}
In this section, the experimental setup is briefly presented. The instrumentation used for the measurement exercises are not cited in detail
as they are standard. The principal instruments used are shown in Figures \ref{1} and \ref{2} and include accelerometers on the beam, an electrodynamic shaker driving the beam through a force transducer and a multichannel signal analyser (Data Physics).\\
The physical  rig consists of a cantilevered  aluminum  beam in contact with an elastic rubber at the free end. The beam was excited at one point with an applied periodic excitation. The beam properties and the rubber stiffness are given in Table $\ref{tab1}$.\\
In practice, the use of a small electrodynamic shaker yields a technical problem due to the reaction of the beam. It is difficult to realize an input force $F(t)$ which is a  simple sine wave unless the impedance of the shaker is significantly higher than the beam impedance . To deal with this problem, a force transducer was fixed between the shaker and the beam to measure the actual supplied excitation force to the beam. The numerical simulations use the actual measured force signal coming from the acquisition system which was periodic. This method is an alternative to modelling the electrodynamic shaker motion. Figure \ref{f1} shows typical examples of  the input force signals with the corresponding spectral content.\\
From a simulation and comparison shown later, using the actual measured force is appropriate given the simplified excitation system without any feedback control which would be necessary to produce a strict harmonic signal.
% \begin{comment}
% \begin{figure}[hbtp]
% \begin{center}
% \includegraphics[width=7cm,height=6cm]{beam_11.eps}%%[width=4cm]
% \includegraphics[width=7cm,height=6cm]{beam_p_contact_11.eps}\label{rig1}
% %\includegraphics[width=3cm,height=5cm]{beam_system.eps}
% \end{center}
% \caption{\small{The rig used for the experiments: a linear clamped-free beam in contact with a rubber support (enlarged photograph on the right)}} \label{1}
% \end{figure}
% \end{comment}

\begin{figure}[hbtp]
\begin{center}
\includegraphics[width=14cm,height=8cm]{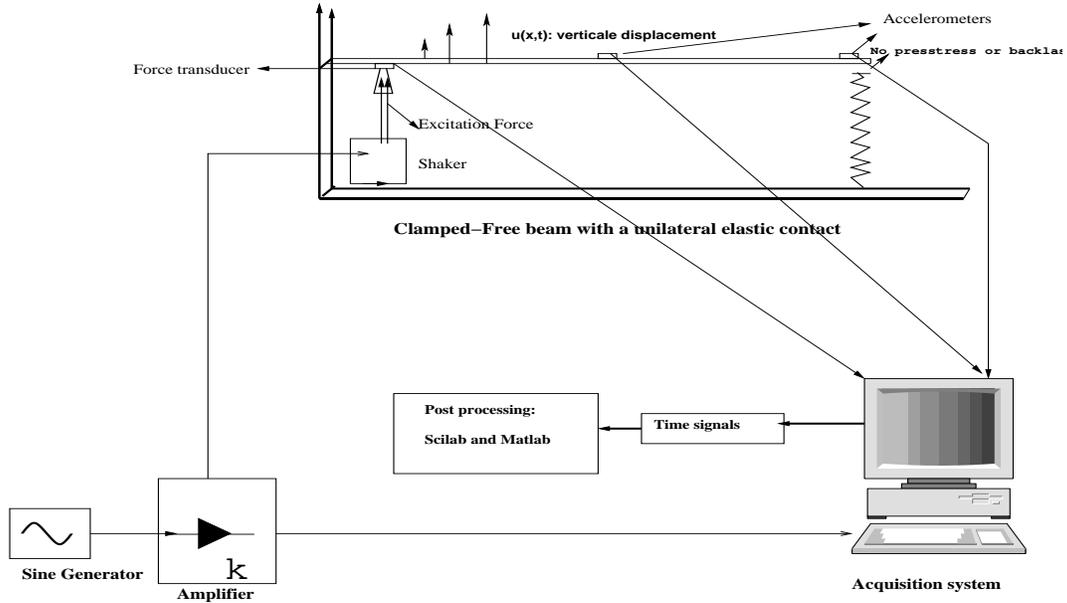}\label{rig2}
\end{center}
\caption{A schematic of the experimental setup.} \label{1}
\end{figure}

\begin{table}[h!b!p!]
\begin{center}
\small{
\begin{tabular}{|c|c|c|c|c|c|}
\hline
& & & & &\\
% after \\: \hline or \cline{col1-col2} \cline{col3-col4} ...
\textbf{Beam} & \textbf{Beam} & \textbf{Beam}& \textbf{Beam Young's}&\textbf{Beam}&\textbf{Spring}\\

\textbf{length} & \textbf{width} &\textbf{ thickness }& \textbf{modulus }& \textbf{density} & \textbf{stiffness} \\
& & & & &\\
\hline
& & & & &\\
$0.35 m $& $0.0385 m$& $0.003m$ & $69\times10^9 N/m^2$&$2700 kg/m^3$&57.14 $KN/m$ \\
%& & & & &\\
\hline
\end{tabular}
\caption{The physical properties of the beam and the spring. The spring stiffness was evaluated using an algorithm described in (\ref{sub42}).}
\label{tab1}}
\end{center}
\end{table}

\section{Beam piecewise linear system dynamics}
The system has two linear configurations or states. The first consists of the beam without the spring in contact and the second when the beam is permanently in contact with the elastic support, which is modeled as a linear spring. The advantage of these two states is to give an idea on the effect of the spring on the overall system dynamics.\\
Usually, adding a spring to  a simple beam model at one point raises the eigen frequency sequence for the system; this shift is realized numerically by adding the spring stiffness  to the coefficient corresponding to the contact point in the stiffness matrix of the system. The spring is massless, so the mass matrix of the F.E. model is intact.\\
Some  experimental problems were encountered; one of these problems is to implement a perfect clamping for the beam, which  is impossible in practice but can be reasonably assumed. Another issue is that the Young's modulus of the rubber spring used for the experiments was unknown. An algorithm was developed to determine the stiffness of the rubber for small displacements. This was subsequently used in the numerical simulations performed for comparison with experiments.
\subsection{The two  linear states}
Firstly, predictions of the forced response of the cantilever beam without a point elastic support were produced and compared with the experiments.
Good agreement showed that the model reasonably accurately  represents the cantilever beam and its boundary conditions. The results for the natural frequencies are shown in Table $\ref{tab2}$. On the other hand, predictions of the cantilever beam with a spring permanently in contact at the free end were produced and compared with the experiments, good agreement was found (see Table $\ref{tab3}$). The last state was used to determine the stiffness of the spring using an algorithm described in the next subsection.
\begin{table}[h!b!p]
\begin{center}
\small{
\begin{tabular}{|c|c|c|c|}
\hline
 & & & \\
 % after \\: \hline or \cline{col1-col2} \cline{col3-col4}
 \textbf{     } & \textbf{ $1^{th}$ natural freq} & \textbf{ $2^{th}$ natural freq}& \textbf{$3^{th}$ natural freq} \\
%% & & & \\
 \hline
 & & & \\
 Predicted& $19.97$ Hz& $122.2$ Hz&$318.8$Hz  \\
 %ù& & & \\
 \hline
 &  &  & \\
Measured & $19.38$ Hz &$118.6$ Hz  & $314.47$ Hz \\
  %%&  &  & \\
 \hline
 & & & \\
 Percentage difference &$3$\% & $3$\% & $1.4$\%  \\
%% &  & & \\
 \hline
\end{tabular}
\caption{The natural frequencies of the clamped-free beam}
\label{tab2}}
\end{center}
\end{table}

\begin{table}[h!b!t!p]
\begin{center}
\small{
\begin{tabular}{|c|c|c|c|}
\hline
& & & \\
 % after \\: \hline or \cline{col1-col2} \cline{col3-col4}
 \textbf{     } & \textbf{ $1^{th}$ natural freq} & \textbf{ $2^{th}$ natural freq }& \textbf{$3^{th}$ natural freq} \\
%%& & & \\
 \hline
& & & \\
 Predicted& $84.57$ Hz& $246.14$ Hz&$443.53$Hz  \\
%%& & & \\
 \hline
& & & \\
Measured & $84.47$ Hz &$243.5$ Hz  & $440$ Hz \\
%%&  &  & \\
 \hline
& & & \\
 Percentage difference &$0.1$\% & $1$\% & $0.7$\%  \\
%%&  & & \\
 \hline
\end{tabular}
\caption{The natural frequencies of the clamped beam with a permanently attached spring}
\label{tab3}}
\end{center}
\end{table}
\subsection{Characterization of the spring support stiffness}\label{sub42}
After initially finding the natural frequencies of the system for the two states of linearity
mentioned in the pervious subsection, an algorithm was developed to find a suitable spring stiffness
by iteration.\\
The shift of the frequencies due to adding a spring at the free end of the beam were recorded. The stiffness matrix of the finite element model incorporated a point spring at the node in contact. Mathematically, the problem is to find a coefficient value $k_r$ which shifts the first eigen frequency $f_0$  of the system without spring to $f_1$, the  first eigen frequency of the same system with a spring permanently in contact. The method is based on the  uniqueness of the sequence of the generalized eigen frequencies of the stiffness and mass matrices.\\
The subsequent value obtained for the point stiffness by this algorithm was $k_r$ equal to $57.14 KN/m$. The corresponding Young's modulus $E_r$, assumes the stiffness $k_r$ equals to the product of the Young's modulus with the spring area divided by the spring length. The estimated Young's modulus for the rubber spring being $4\times 10^6 N/m^2$.
%%\newpage
\section{Comparison of simulations with experiments}
In this section the simulations are compared to the measured data in the frequency domain; the acquisition system provides just the time signals of the accelerations and the input force. The processing of these signals and the numerical results were done using external software (Scilab \cite{a6}).\\
The  total length of data predicted corresponds to an integration time which is fixed at $t$ equal to $1s$ for all of the simulations and the acquired experiment of samples. It is five times the fundamental (lowest) period of the system. The data are measured immediately above the support and the frequency axis is normalized by the excitation frequency\\
For the industrial application it was necessary to consider the response in the  first three modes, hence the beam was modelled using ten equal length  finite elements. In principle, the model is able to be applied to higher frequency excitations but typically any fatigue or damage in practice is likely to occur in the lower order modes.
\subsection{Comparison in the frequency domain}
The effect of the unilateral contact is clear, the input frequency is split into its all harmonics. From an energetic point of view, the input energy is split, each subharmonic of the main excitation takes its part thus the contribution of each harmonic is  evident.\\
Figures \ref{fft_32}, \ref{fft_124} and \ref{fft_100} show the FFT of the numerical and the experimental displacements for an excitation signal at $32$ Hz, $124$ Hz and $100$ Hz respectively. The height of the peaks are normalized by the maximum. The predicted frequencies found are exactly the same as measured for a large number of harmonics. However, a small shift in the height of these peaks appears for the fifth harmonic; the peak in  Figure \ref{fft_32}  appears at a multiple of 5 times the original main excitation frequency, i.e. at approximately 160Hz in the  acceleration response.  At this frequency there is no guarantee that the actual support of the beam and the spring is itself rigid, as it might have its own dynamics as would the bench that supports the rig, so there might be some influence of that on the response. Other tests with random excitations have shown good agreement.
Figures  \ref{f1} and  \ref{f2} show the input excitation force and the measured acceleration at $32$ Hz and $124$ Hz respectively. It is clear from the time signal and from the frequency content that the forces are not pure harmonic single frequency sine waves, but they are periodic.\\
Figures \ref{d_32}, \ref{d_124} and \ref{d_100} show the predicted displacement for an excitation at $32$ Hz, $124$ Hz and $100$ Hz respectively. The displacements are almost always positive so they have  positive means. This is due to the high stiffness of the spring, but the time response is still periodic.\\
\subsection{Magnitude-Energy dependence}
The magnitude-energy dependence is a typical dynamical feature of nonlinear systems under  excitation; the maximum of the solution
plotted against the input energy can take different shapes depending on the  form of the nonlinearity.
For linear systems under periodic excitation, the maximum of the solution is proportional to the input energy. The model studied in this paper is a piecewise linear system, the numerical and the experimental  results is expected to exhibit a linear behaviour for different levels of input energy.\\
The magnitude-energy dependence can be represented by different ways. Herein, a mathematical and an experimental proof are presented to demonstrate the magnitude-energy independence. From a mathematical point of view, the level of excitation energy depends on the magnitude of the excitation force $F(t)$. The idea here is to examine the variations of the solution in the time domain as the amplitude of the excitation force $F(t)$ changes linearly. Consider then equation $(\ref {equation1})$ and  multiply both sides by a constant $\lambda\geq0$ the equation becomes:
\begin{equation}
\lambda[\rho S\ddot{u}(x,t)+EIu^{(iv)}(x,t)]=\lambda[F(t)\delta_{x_0}-(k_r u(x_1,t)_-)\delta_{x_1}] \label{equation2}
\end{equation}
In general, the only problem to substitute the parameter $\lambda$ in the equation is the nonlinear term; in this case, $\lambda[k_r u(x_1,t)_-]=k_r [\lambda u(x_1,t)]_-$ (see definition of $u_-$). Equation $(\ref{equation2})$ can then be written as follow:
\begin{equation}
\rho S\ddot{v}(x,t)+EIv^{(iv)}(x,t)=\lambda [F(t)\delta_{x_0}]-k_r (v(x_1,t)_-)\delta_{x_1} \label{equation3}
\end{equation}
such that $v=\lambda u$. In conclusion, the solution $v$ of the PDE governing the motion is proportional to the excitation force $F(t)$.\\
Note that this substitution for the parameter $\lambda$ is not  generally possible, e.g. if a prestress is applied between the spring and the beam; it is the case
for many other nonlinearities too ($\lambda x^3\neq(\lambda x)^3$).\\
Experimentally, the mean square responses of each harmonic is plotted against the power spectral density $G_{xx}$ of the input force for three
levels of excitation. The mean square response of the mode is calculated approximately by using the Mean Square Bandwidth $\pi \zeta \omega_n$. The experimental estimate of the equivalent viscous damping ratio is  $\zeta=\frac{\omega_2-\omega_1}{2\omega_n}$, where $\omega_n$ is the resonance frequency, $\omega_1$ and $\omega_2$ are the frequencies corresponding to the half-power points (-3dB below the maximum peak response).\\
Usually, this method is used to approximate the mean square response at the natural frequency of a linear system. Herein, it is used at the first resonance frequency of the nonlinear system ($32$ Hz) which can be calculated as the inverse of the mean of the linear periods of the two piecewise linear systems; it is also applied to its harmonics.
% The concept of the mean square response can be applied to the fundamental mode of the nonlinear system and for each harmonics.
Figure \ref{ed} shows the mean square responses for the fundamental mode and for the first two harmonics normalized by the excitation mean square level, against three different input levels. As the excitation level increases the response at the excitation frequency and its harmonics increases proportionally, the relationship between the fundamental mode and its harmonics is linear; this linear behaviour is due to the linearity of the spring and the beam.

\section{Numerical simulations}
In this section, some further numerical simulations are presented in order to investigate and  understand better the system behaviour. Figures \ref{d_sine_32} and \ref{d_sine_124}  show the displacement for a sine excitation at $32$ Hz and $124$ Hz respectively; Figure \ref{fft_sine_32} shows the frequency content of the displacement for a sine excitation at $32$ Hz; the results cannot be compared to the experiments as it is not easy to realize a simple sine excitation. The dynamic behaviour of the system is similar to those presented in the previous section. The main response is split into all the subharmonics of the excitation frequency.\\
The elastic force from the spring  is applied to the free end of the beam and should be taken in account as it can damage the structure; this force is non differentiable as the spring is only on contact when the beam has a negative displacement and its magnitude is proportional to the spring compression. Figure \ref{force_sine_32} and \ref{force_sine_124} show the predicted time signal of the force applied to the beam for a sine excitation at $32$ Hz and $124$ Hz respectively. Note that in case of a bilateral spring (spring attached to the beam), the time signal should be differentiable  and periodic
with a zero mean.
%%\newpage
\section{The effect of the unilateral spring position}
% \begin{comment}
% \begin{figure}[hbtp]
% \begin{center}
% \includegraphics[width=14cm,height=6cm]{beam_system_3.eps}
% \end{center}
% \caption{\small{beam system with an unilateral spring under a periodic excitation}} \label{2}
% \end{figure}
% \end{comment}
An aim of this work is to provide a model which can predict the dynamic behaviour of a beam striking an elastic support.
It is also necessary to choose the preferable points of the structure to position the support. In this section, the spring is moved to the node of the second linear beam  mode  which corresponds to  a particular node of the F.E. model (see Figure \ref{2}). The subsequent numerical  results
show a linear behaviour of the system for an excitation near the second eigen frequency. They  show
a nonlinear behaviour as in the previous case for any other frequency of  excitation. The results are presented in the
frequency domain as before taking in account the new spring's position. Figure \ref{fft_new_122} shows the FFT of the numerical and the experimental displacements for an excitation at $122$ Hz, very close to the second eigen frequency of
the linearized system (see Table \ref{tab2}). It is clear that the response primarily has a single frequency content as the input signal (Figure \ref{force_new}), which is a fundamental property of a linear system.\\
Figure \ref{fft_new_32} shows the FFT of the experimental and the numerical displacements for an excitation of $32$ Hz.
The input frequency is split into all subharmonics, the behaviour of the beam is the same as described in  the previous section, as the system is no longer linear.
\section*{Conclusions}
A numerical and experimental study of a beam with a unilateral elastic contact has been presented, the model used
for the predictions having been validated by experiments. The comparison was performed in the frequency domain for different excitation frequencies; the results showed a very good agreement.\\
The comparison in the time domain needs a sophisticated processing of the time domain signals to eliminate or reduce the contribution from higher order frequencies not involved in the motion; this aspect will be in the scope of the future.\\
The results showed the effect of the spring position on the dynamic behaviour; other positions could be of interest if the system is subjected to high frequency excitation as the number of nodes increase with respect to the excited modes. Some experimental results for a pres-stressed contact is currently  under investigation, this will be reported in the future. Also, future work will consider other types of excitation such as broadband random base excitation which might be present for the practical application of launching stacked solar array panels.
%%\newpage
\section*{Acknowledgements}
This work was conducted for a  PhD project of the first author with a scholarship from Thales Alenia Space, France. The authors also gratefully acknowledge the financial support of the ``Conseil G\'en\'erale des Alpes Maritimes'' to realize this project.
\bibliographystyle{unsrt}
\bibliography{biblio}
%%%%%%%%%%%%%%%%%%%%%%%%%%%%%
\begin{figure}[hbtp]\label{solar}
\begin{center}
\includegraphics[width=6cm,height=6.5cm]{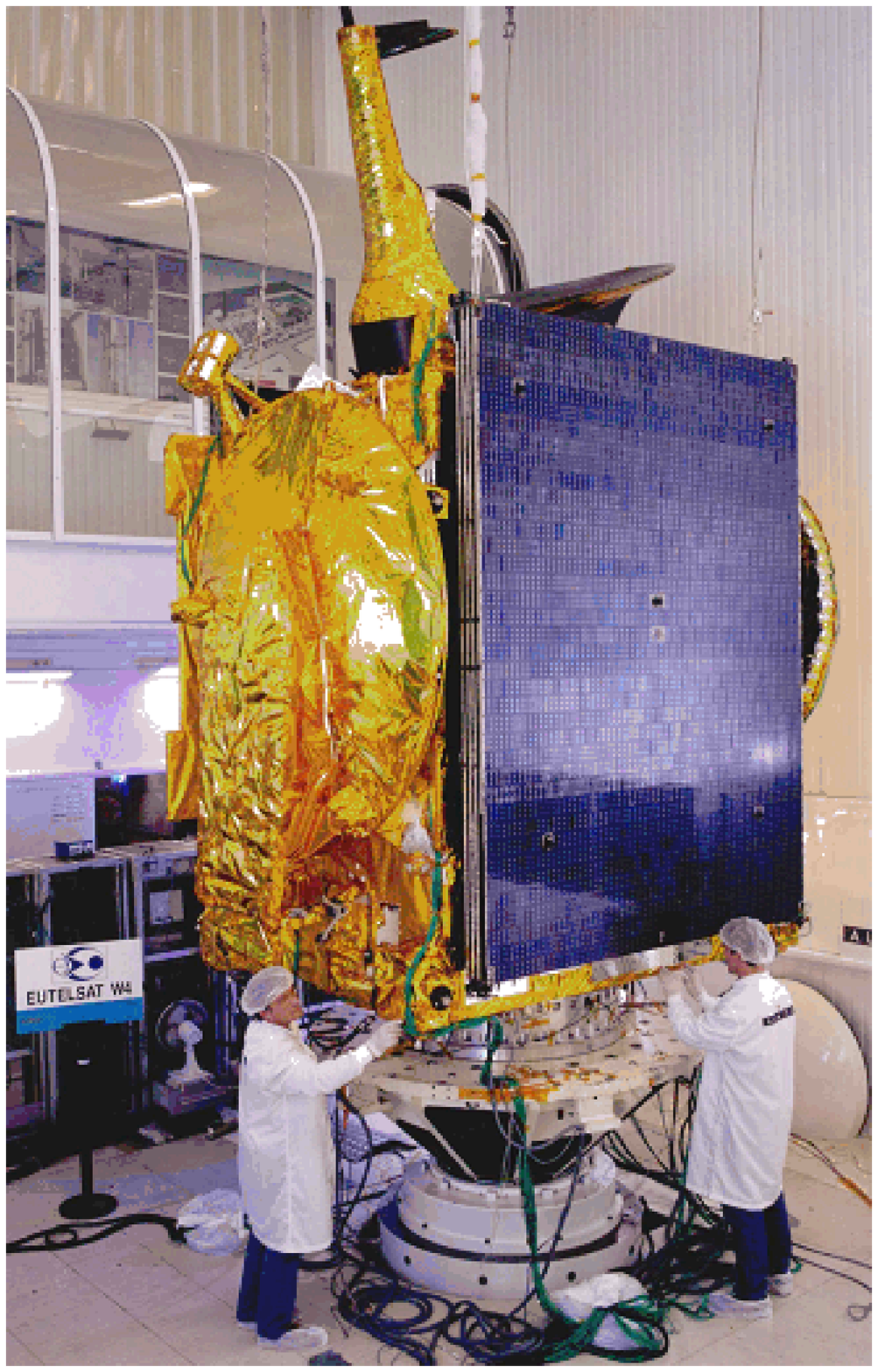}%%[width=4cm]
\hspace{0.1cm}
\includegraphics[width=6.5cm,height=6.5cm]{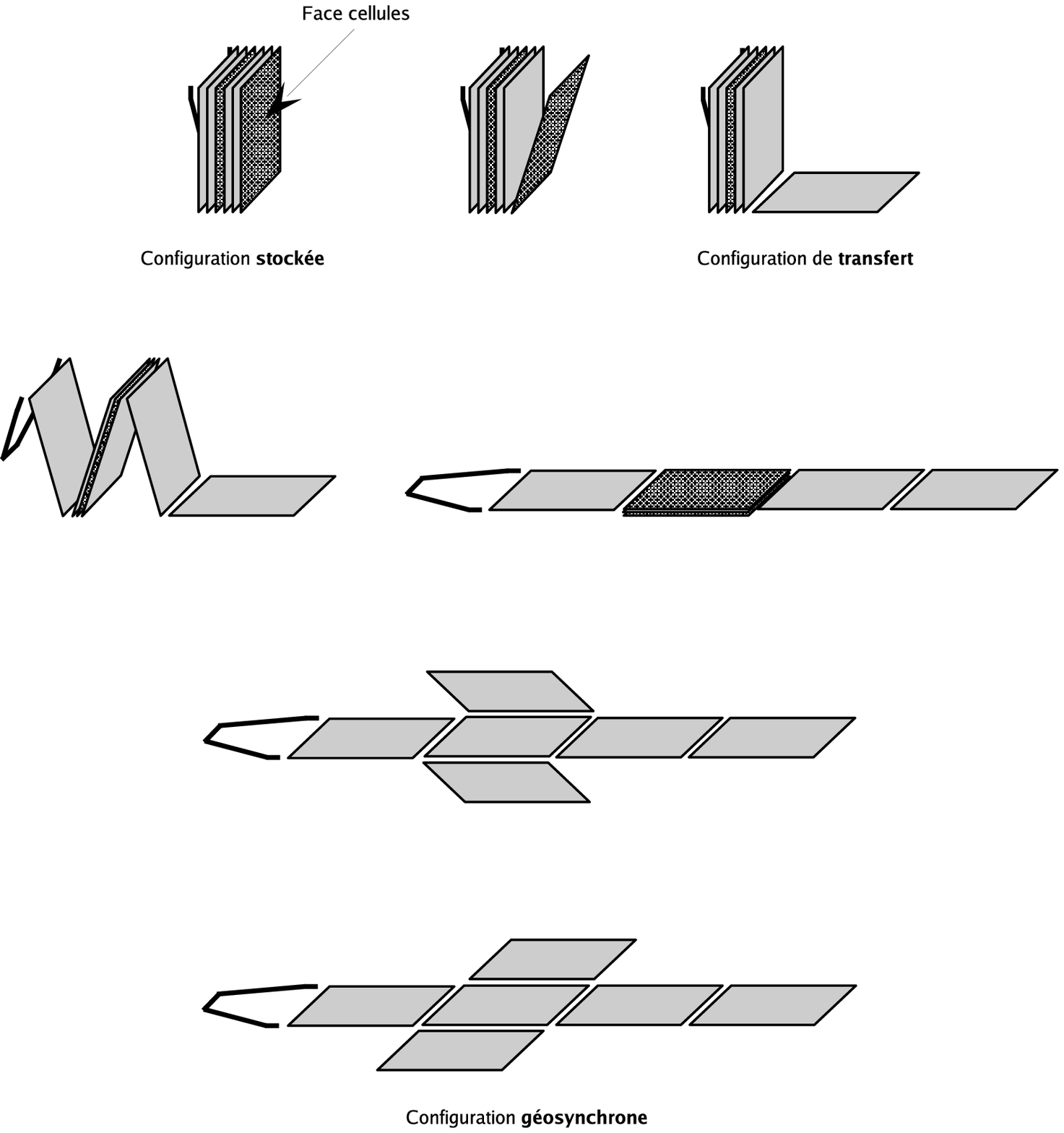}
%\label{solar1}
\end{center}
\caption{\small{Left: Solar array of a satellite under a test on a shaker. Right: A solar array from the folded to the final position}}
\end{figure}
%%%%%%%%%%%%%%%%%%%%%%%%%%%%%%%%%%%%%%%%%%%%%%%%%%%%%%%%%%%%%%%%%
\begin{figure}[hbtp]
\begin{center}
\includegraphics[width=14cm,height=7cm]{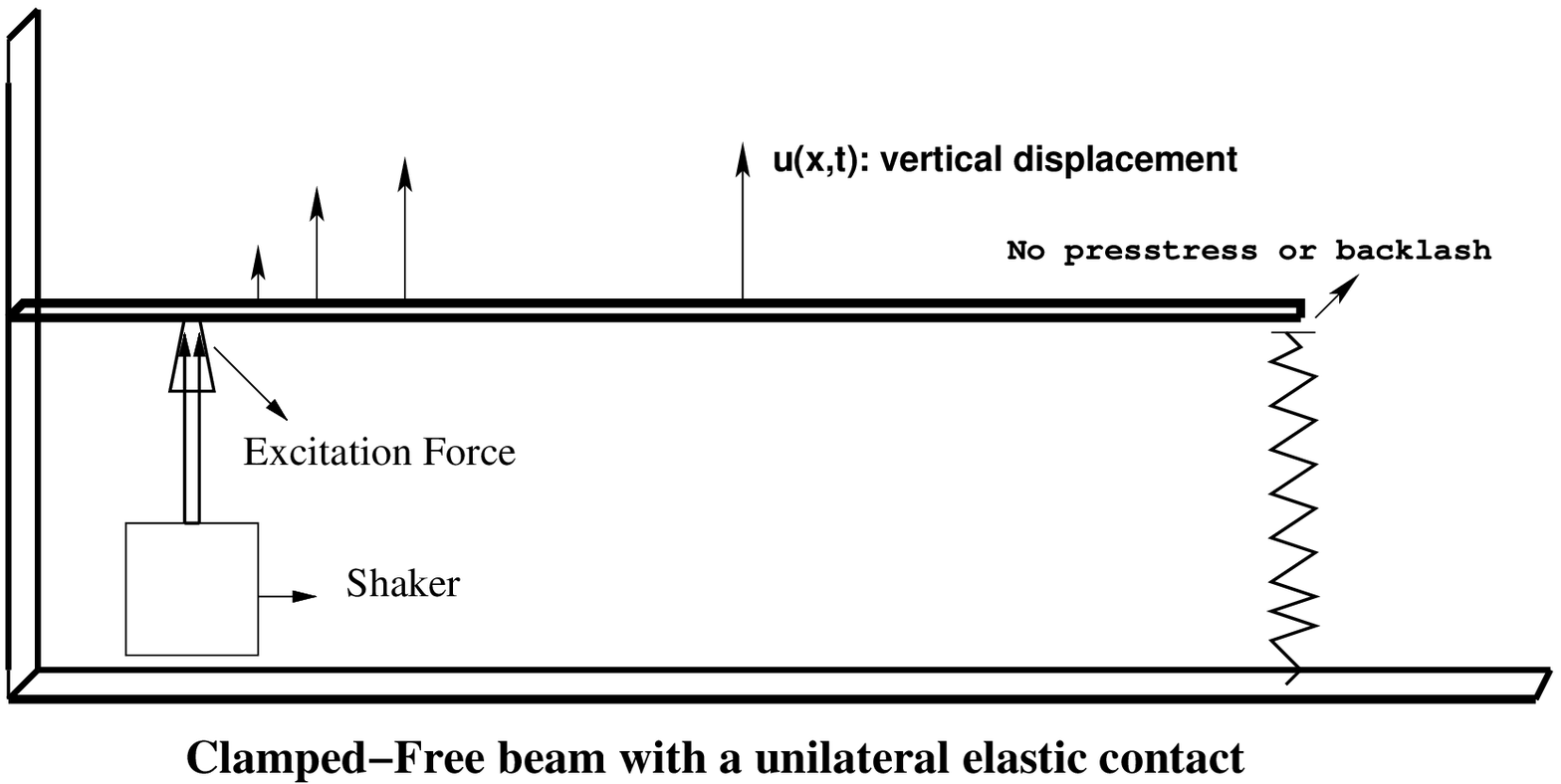}
\end{center}
\caption{\small{beam system with an unilateral spring under a periodic excitation}}
\label{1}
\end{figure}
%%%%%%%%%%%%%%%%%%%%%%%%%%%%%%%%%%%%%%%%%%%%%%%%%%%%%%%%%%%%%%%%%
\begin{figure}[hbtp]
\begin{center}
\includegraphics[width=7cm,height=6cm]{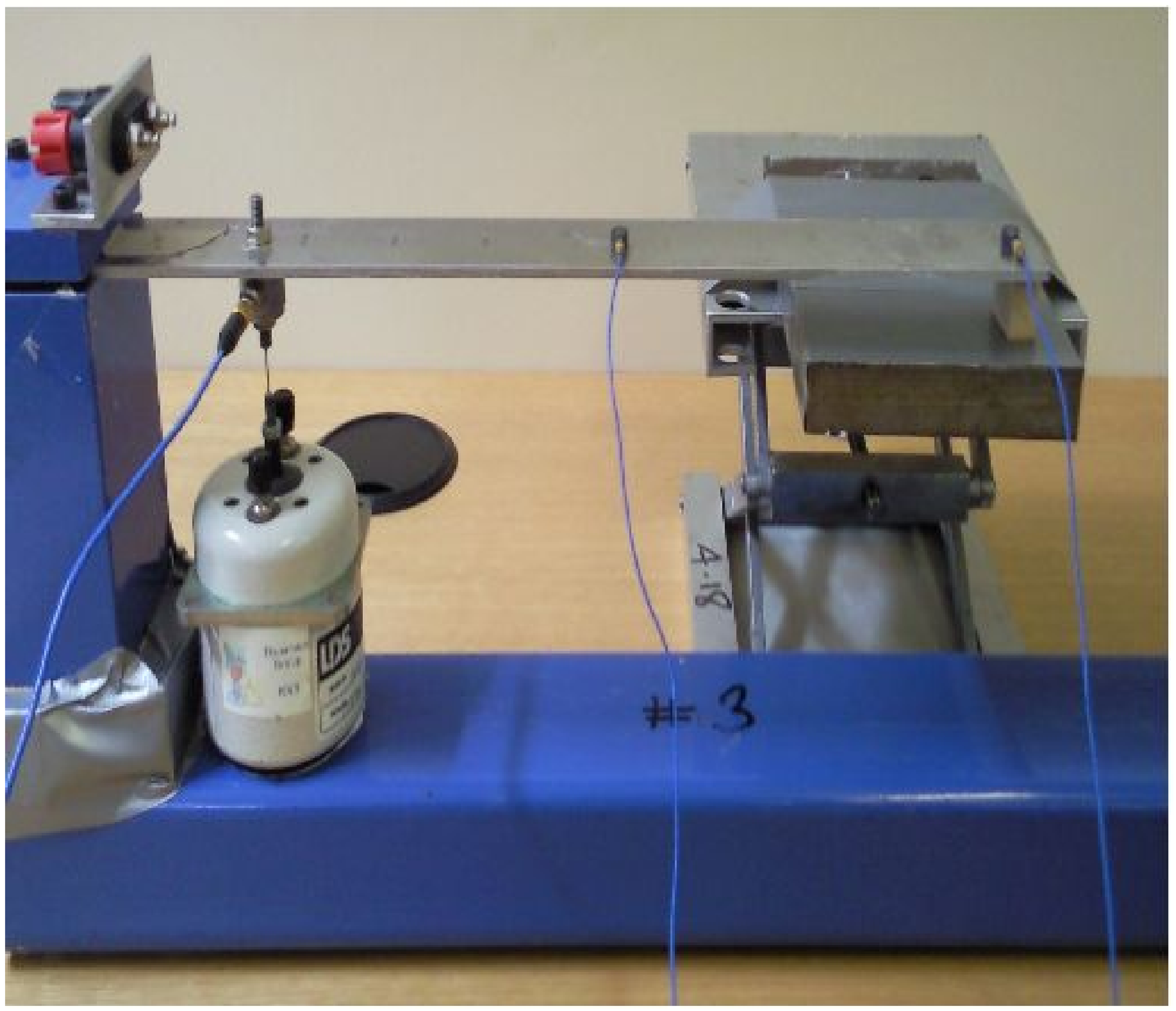}%%[width=4cm]
\includegraphics[width=7cm,height=6cm]{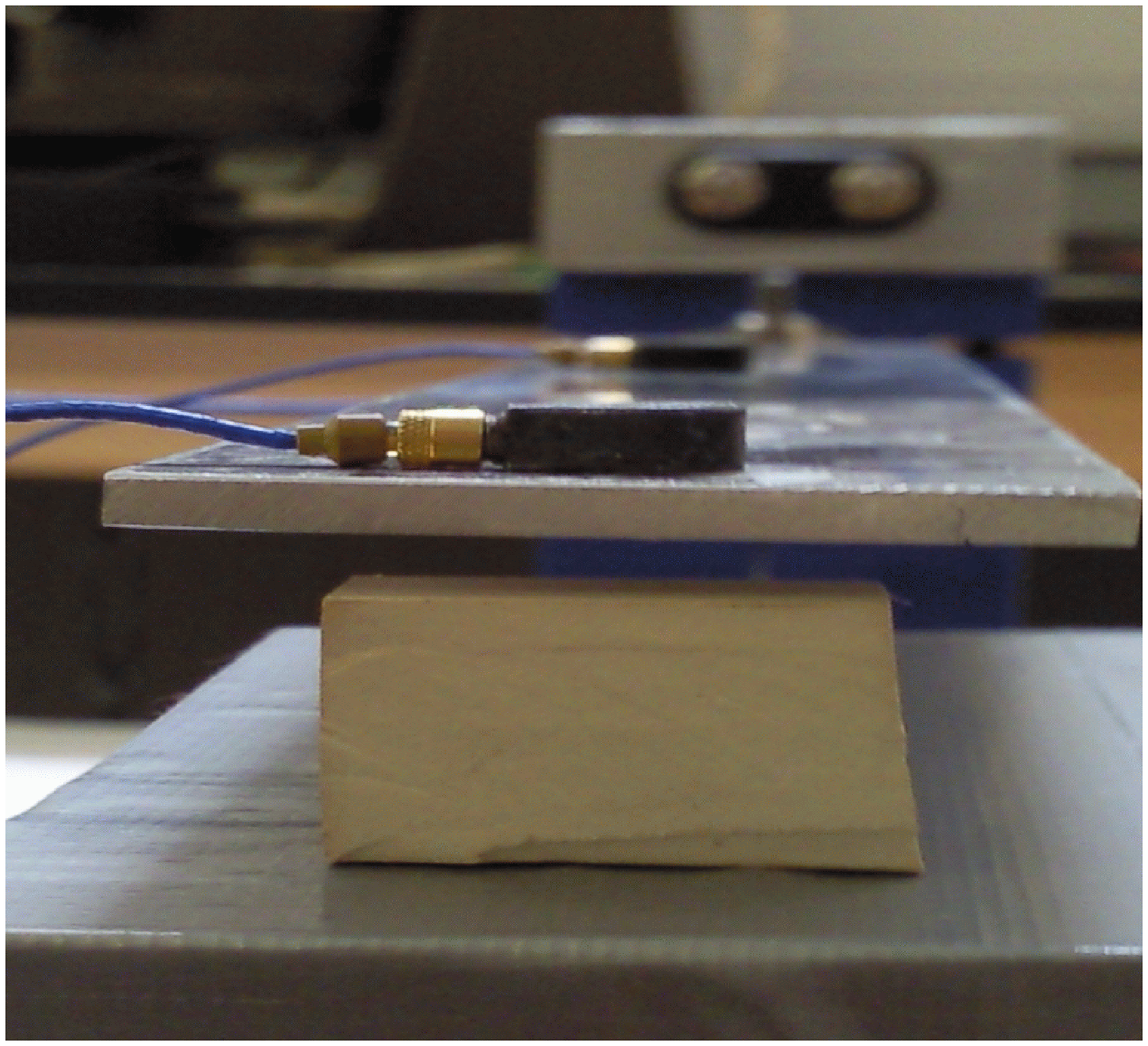}\label{rig1}
\end{center}
\caption{\small{The rig used for the experiments: a linear clamped-free beam in contact with a rubber support (enlarged photograph on the right).}} \label{2}
\end{figure}

\begin{figure}[hbtp]
\begin{center}
\includegraphics[width=14cm,height=6.5cm]{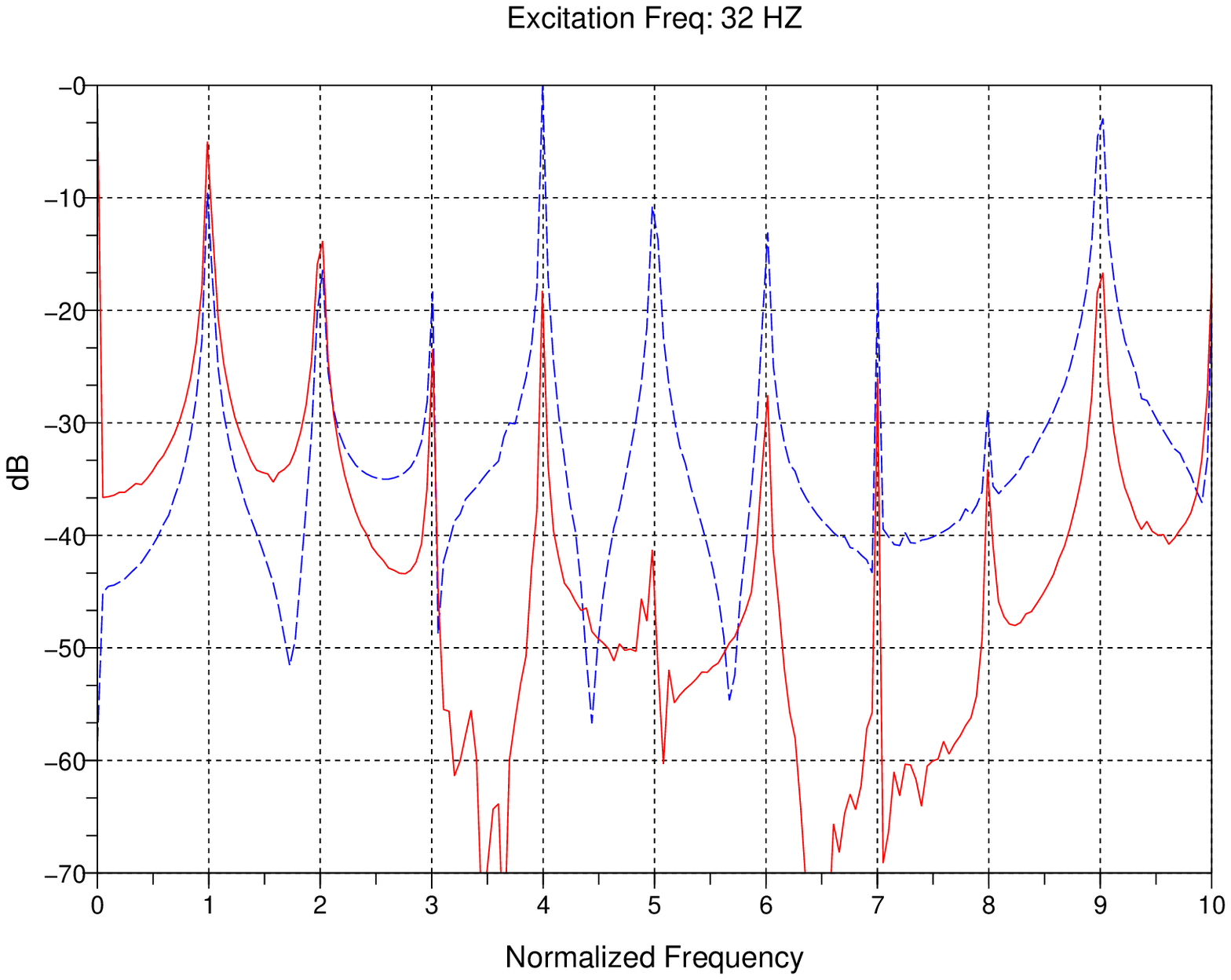}
\caption{Predicted (solid) and measured displacements (dashed) (dB) for an excitation at $32$ Hz applied to the beam with unilateral support stiffness. The displacement is normalized by the peak value and is measured immediately above the support and the frequency axis is normalized by the excitation frequency.}
\label{fft_32}
\end{center}
\end{figure}
\begin{figure}[hbtp]
\begin{center}
\includegraphics[width=14cm,height=6.5cm]{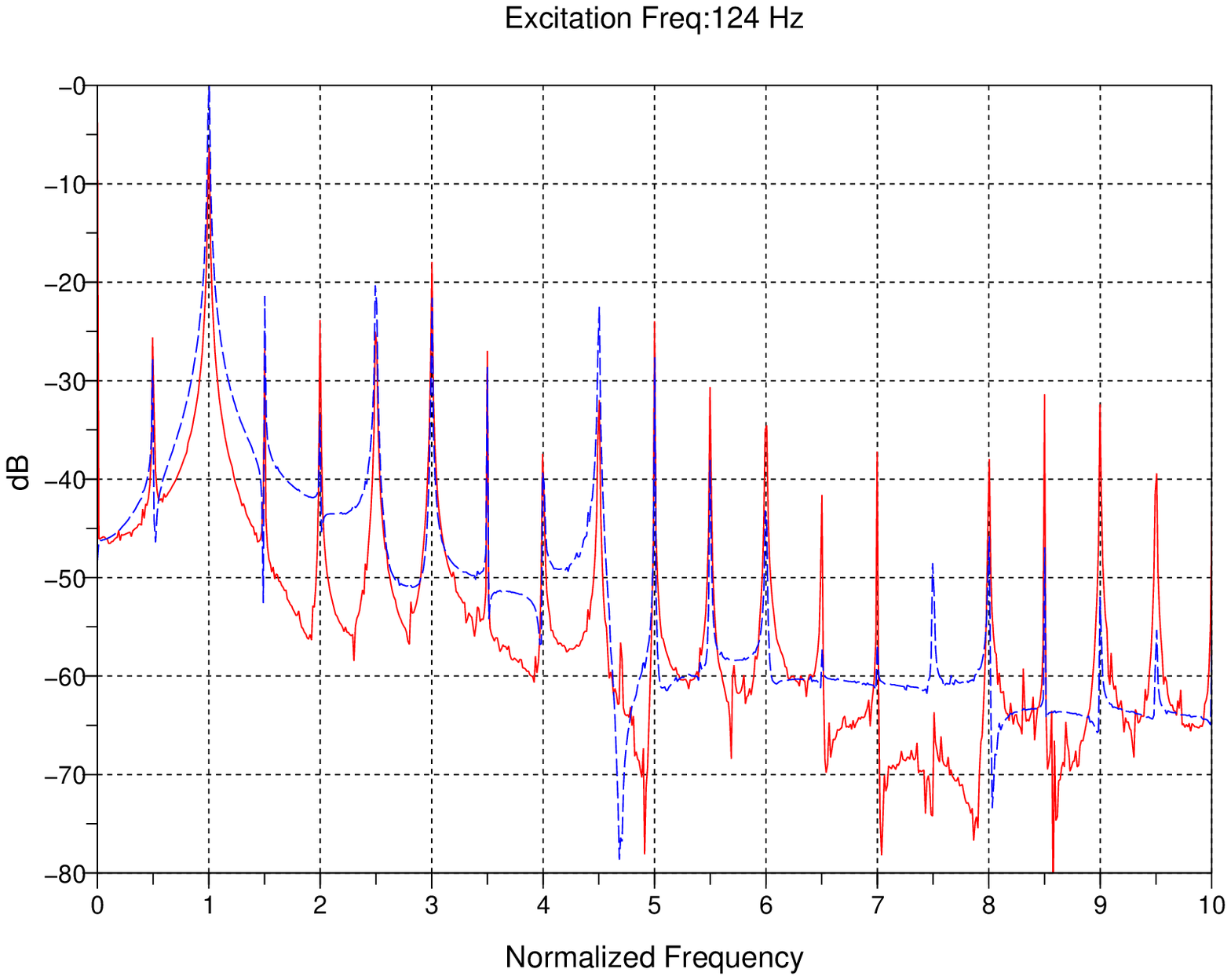}
\caption{Predicted (solid) and measured displacements (dashed) (dB) for an excitation at $124$ Hz applied to the beam with unilateral support stiffness. The displacement
is normalized by the peak value and is measured immediately above the support and the frequency axis is normalized by the excitation frequency.}
\label{fft_124}
\end{center}
\end{figure}

\begin{figure}[hbtp]
\begin{center}
\includegraphics[width=14cm,height=7cm]{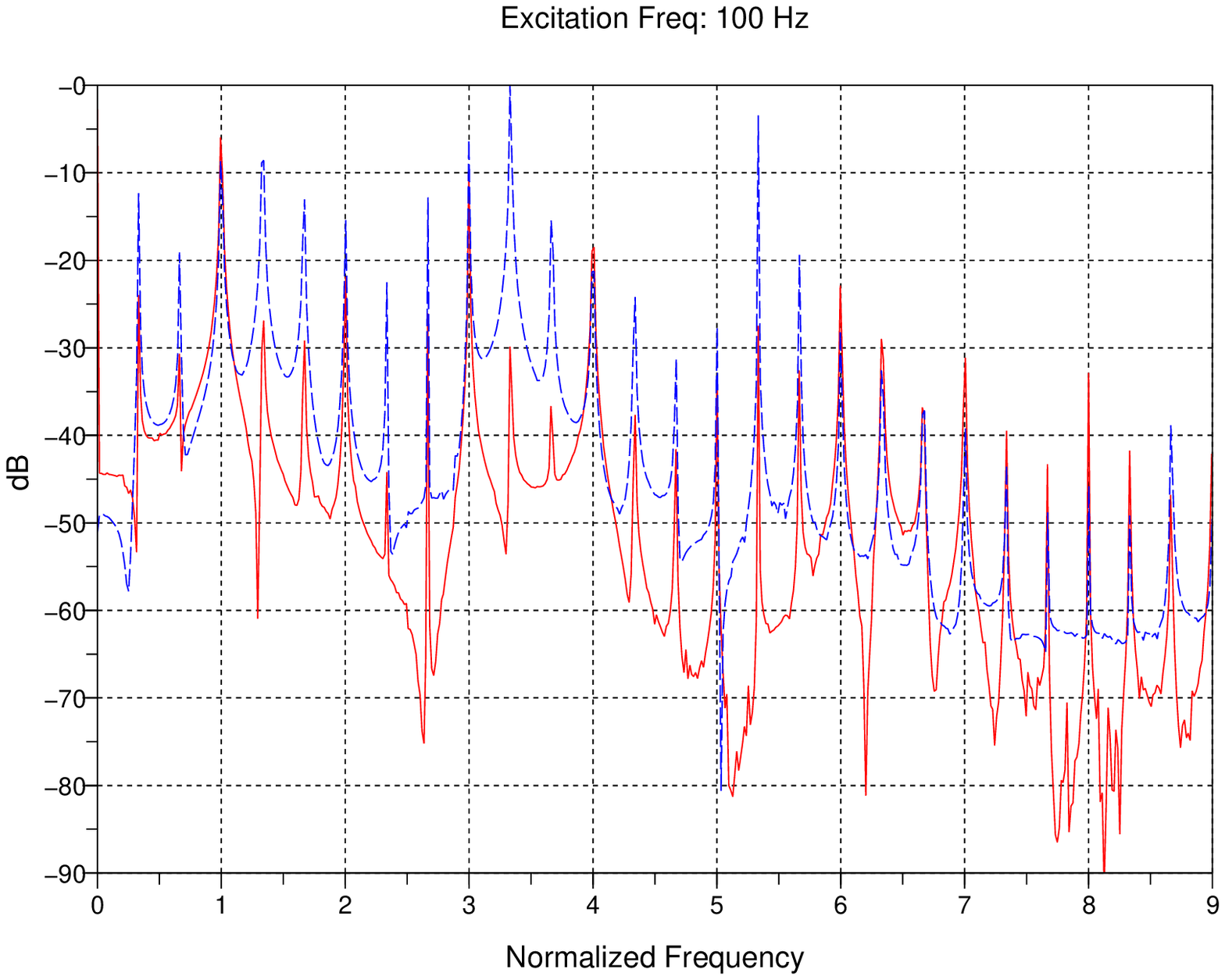}
\caption{Predicted (solid) and measured displacements (dashed) (dB) for an excitation at $100$ Hz applied to the beam with unilateral support stiffness. The displacement
is normalized by the peak value and is measured immediately above the support and the frequency axis is normalized by the excitation frequency.}
\label{fft_100}
\end{center}
\end{figure}
\begin{figure}[hbtp]
%%%%%%%%%%%%%%%%%%%%%%%%%%%%%%%%%%%%%%%%%%%
\begin{center}
\includegraphics[width=14cm,height=4.5cm]{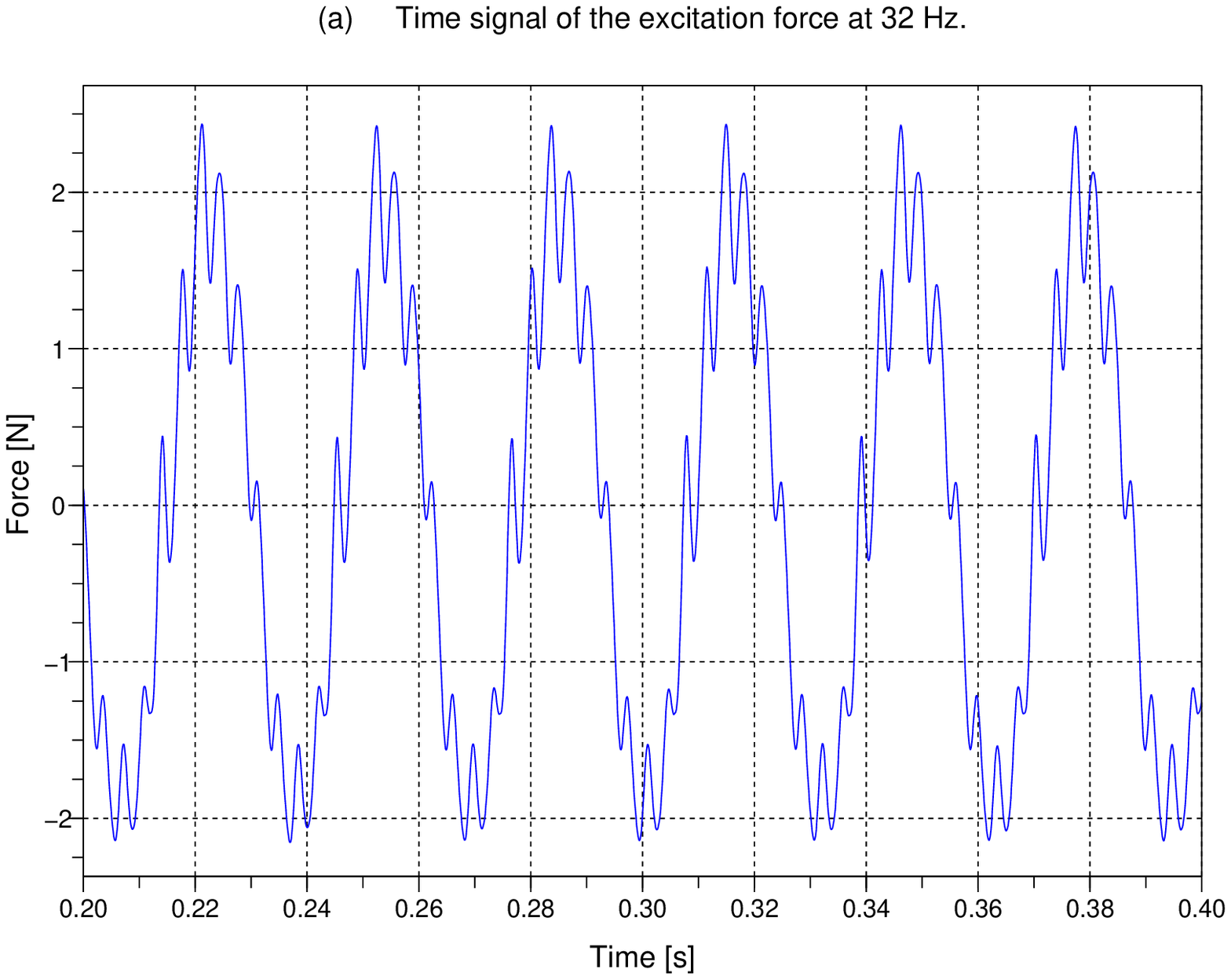}
\vspace{0.5cm}
\includegraphics[width=14cm,height=4.5cm]{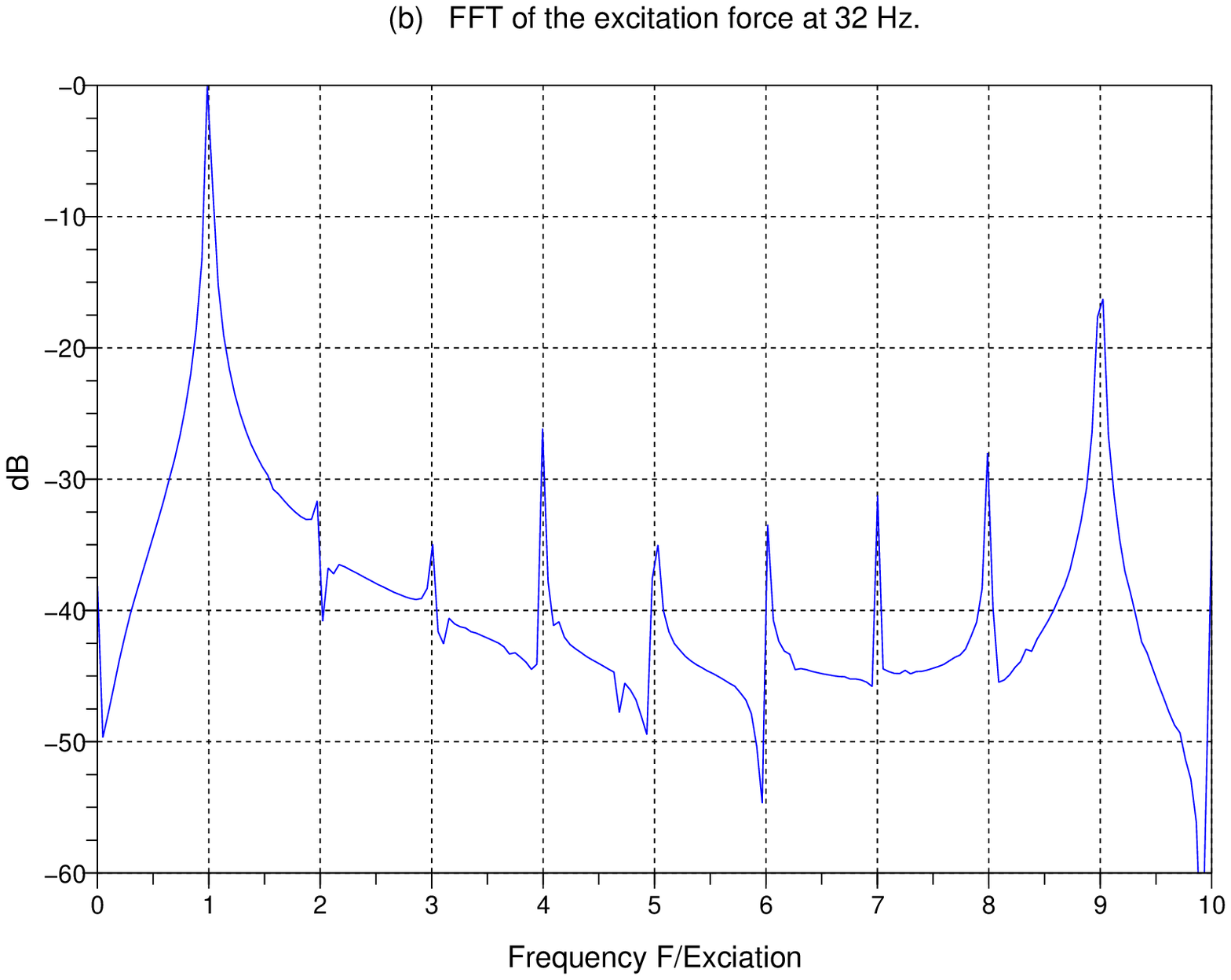}
\includegraphics[width=14cm,height=4.5cm]{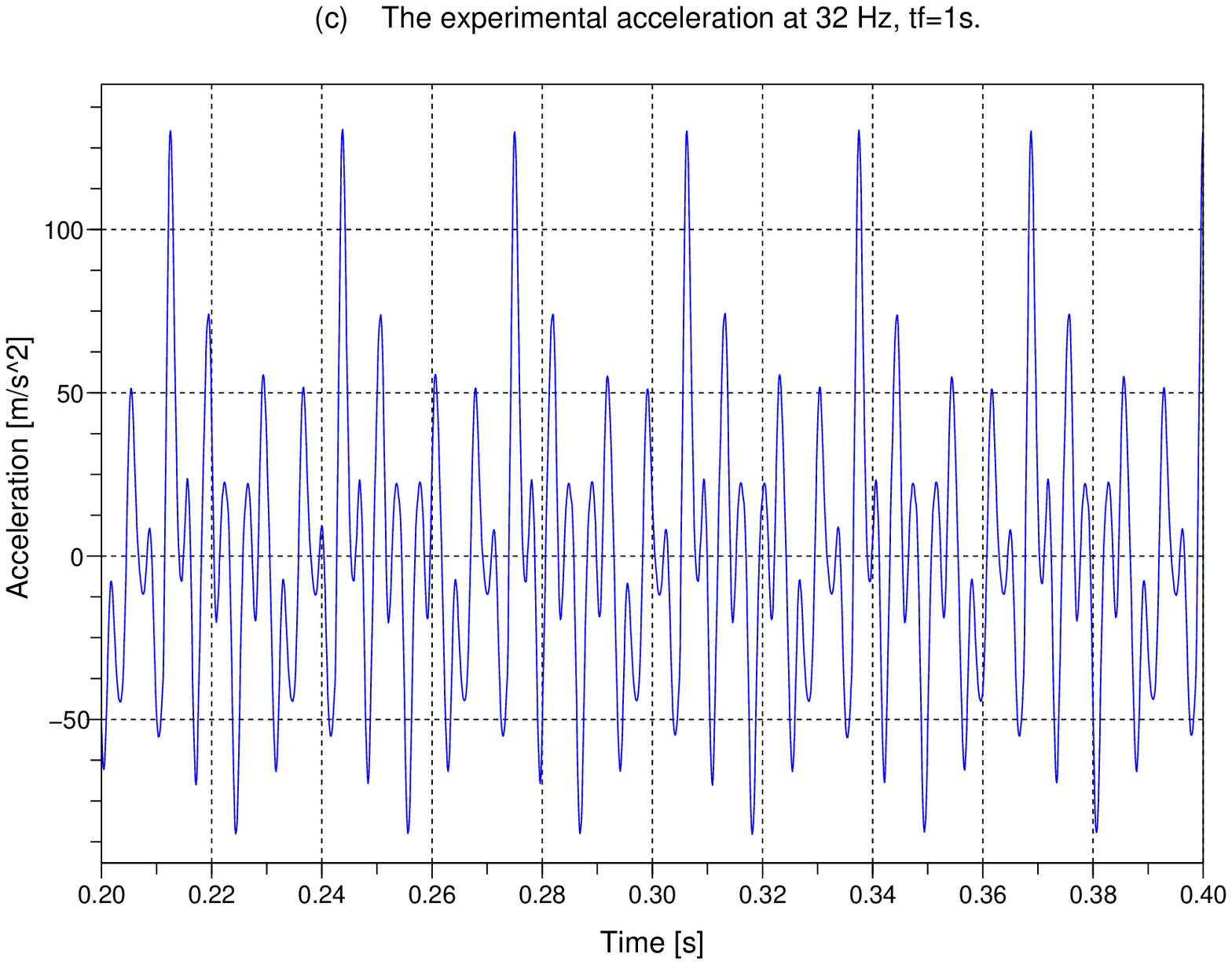}
\includegraphics[width=14cm,height=4.5cm]{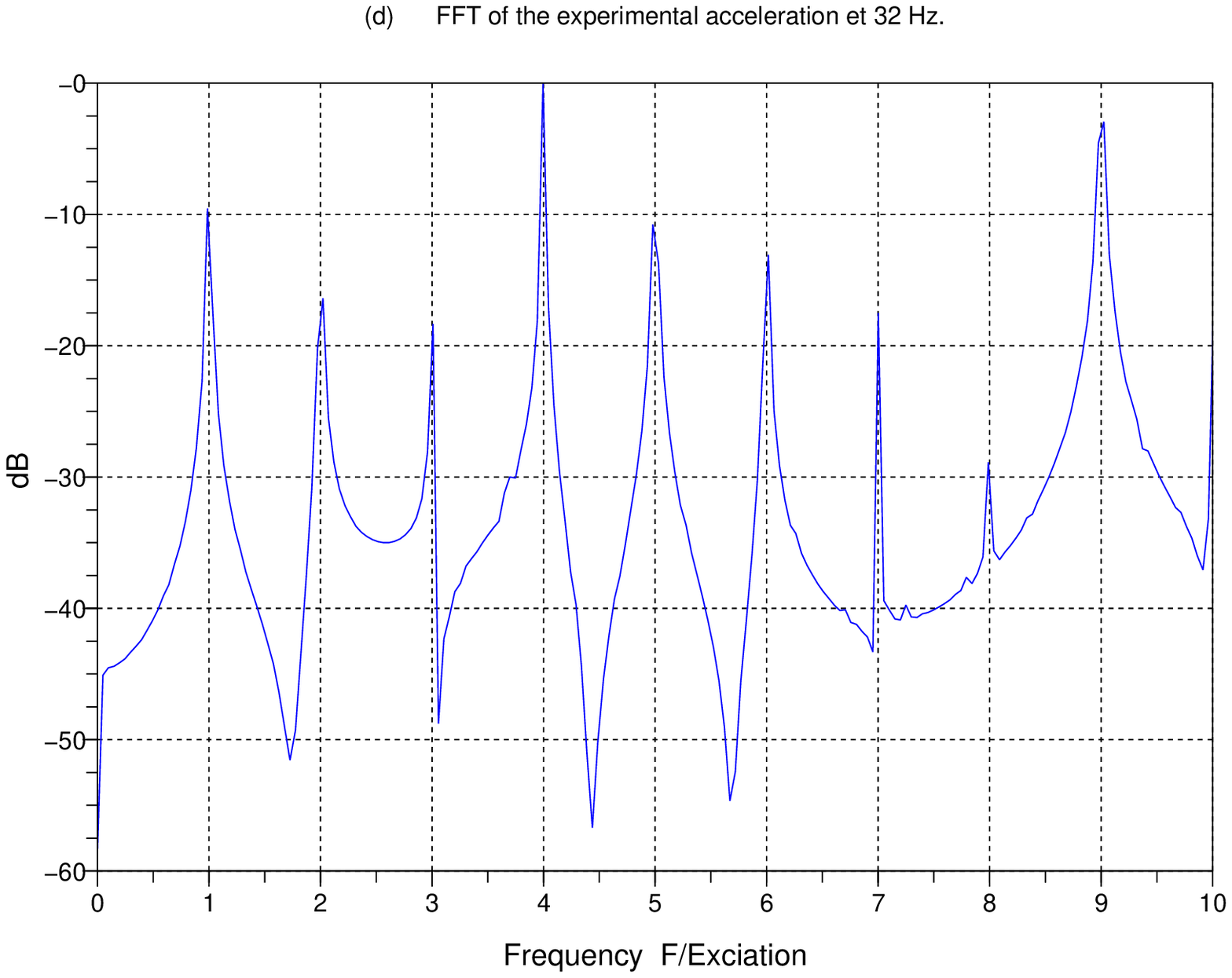}
\caption{Measured excitation force (a) and its frequency contents (b),  measured acceleration response (c) and its frequency content (d) for an excitation at $32$ Hz. Strictly the force is not harmonic.}
\label{f1}
\end{center}
\end{figure}
\begin{figure}[hbtp]
\begin{center}
\includegraphics[width=14cm,height=4.5cm]{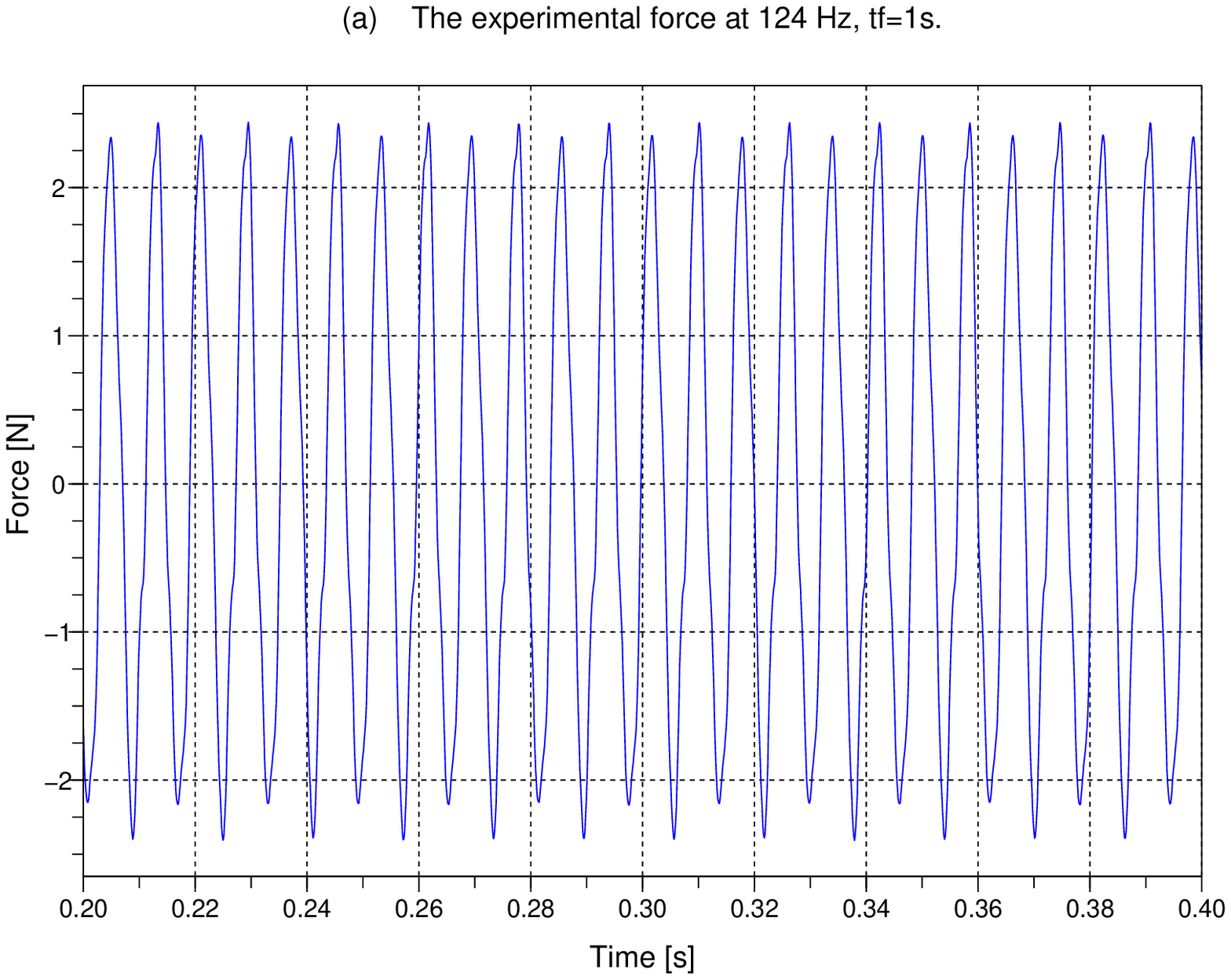}
\vspace{0.5cm}
\includegraphics[width=14cm,height=4.5cm]{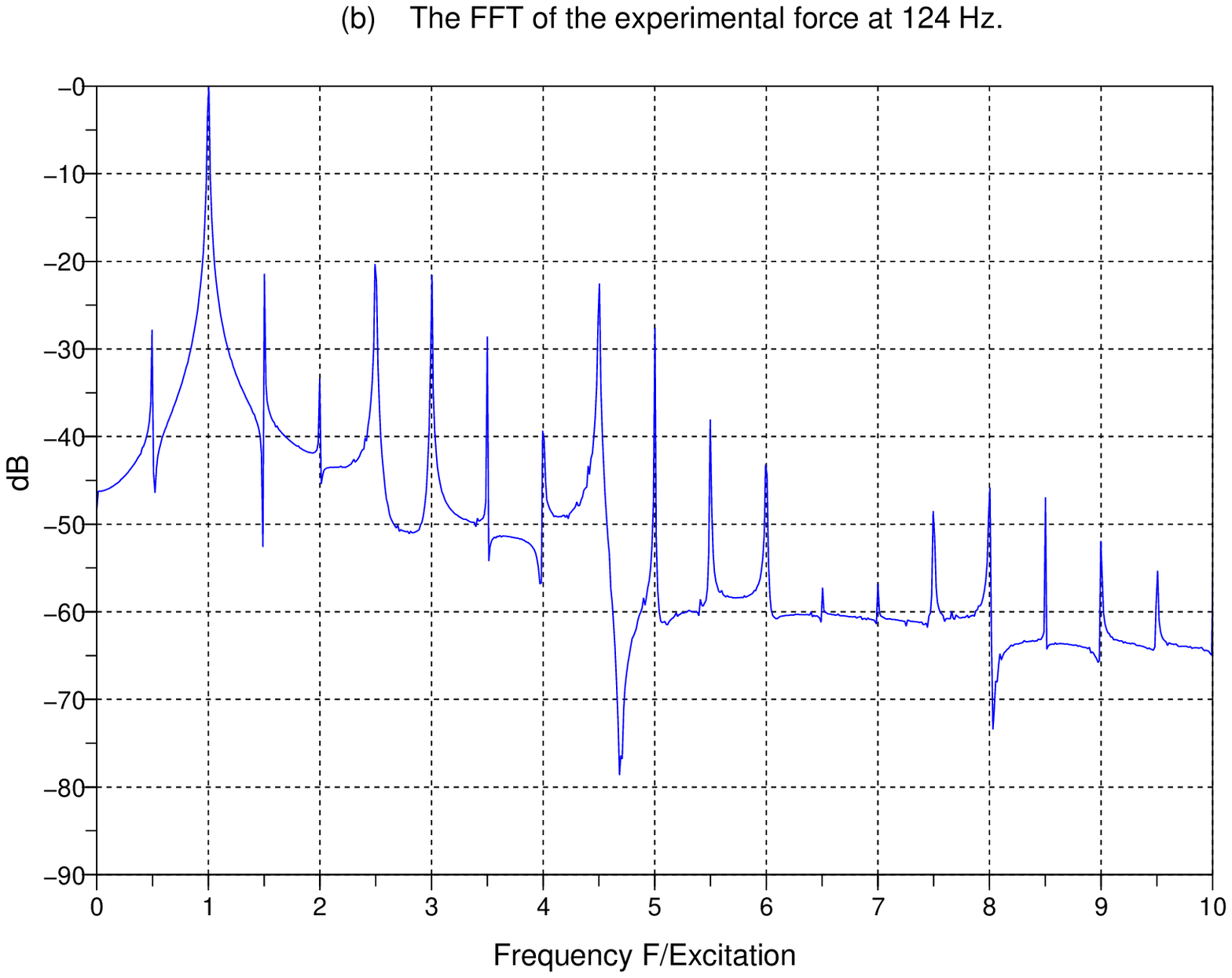}
\includegraphics[width=14cm,height=4.5cm]{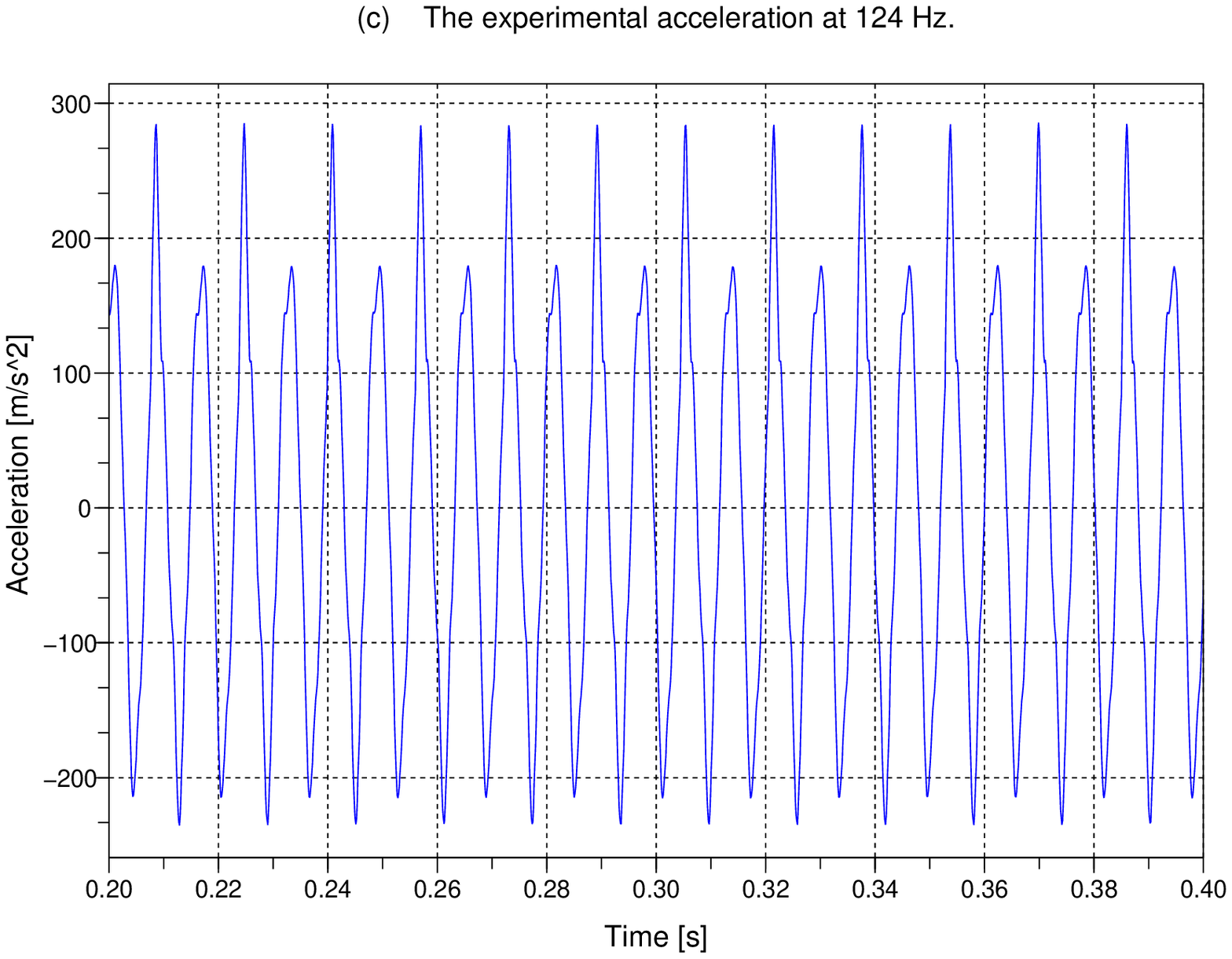}
\includegraphics[width=14cm,height=4.5cm]{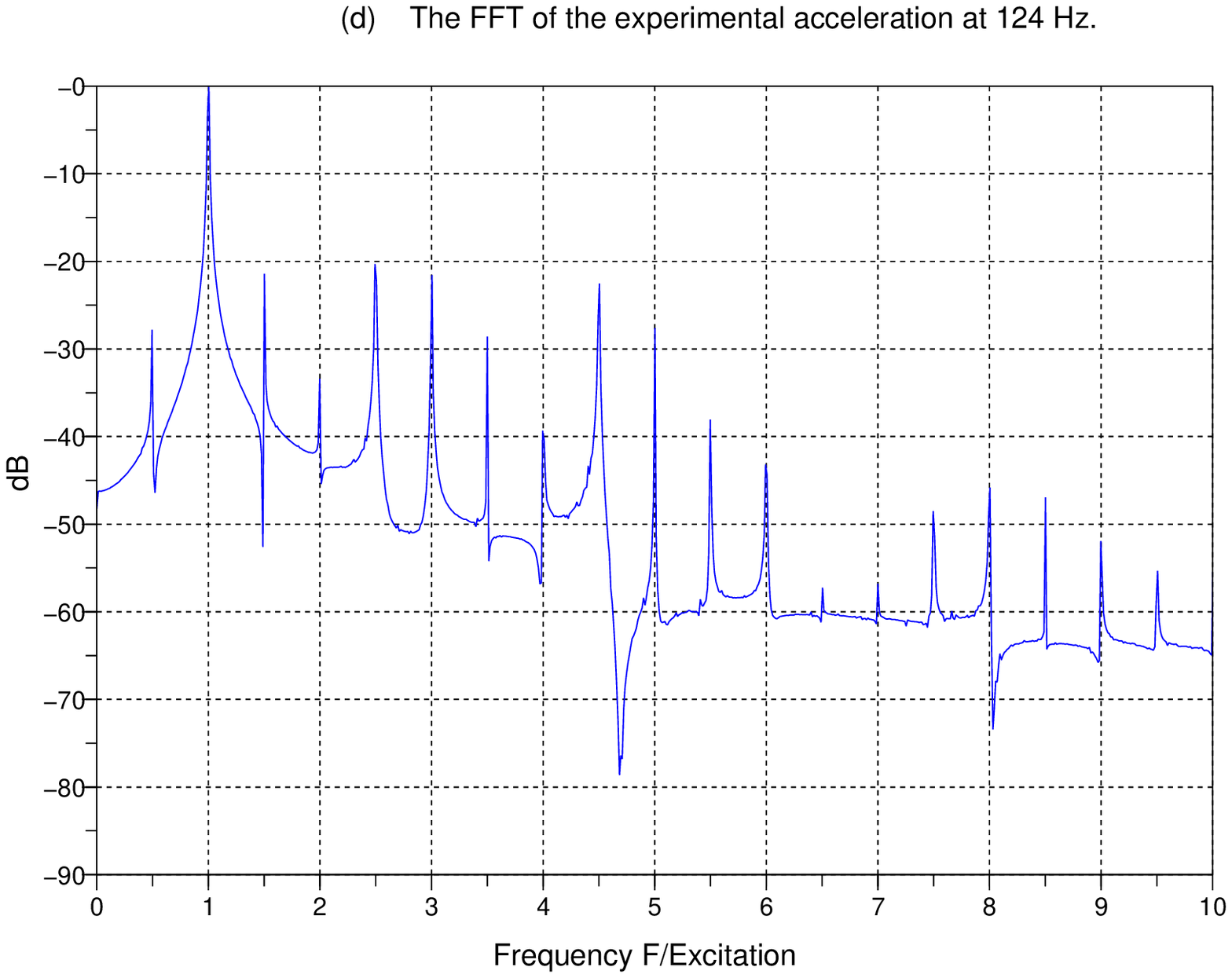}
\caption{Measured excitation force (a) and its frequency contents (b), the measured acceleration response (c) and its frequency content (d) for an excitation at $124$ Hz. Strictly the force is not harmonic.}
\label{f2}
\end{center}
\end{figure}
%%%%%%%%%%%%%%%%%%%%%%%%%%%%%%%%%%%%%%%%%%%%%%%%%%%%%%%%%%%%%%%%%%%%%%%%%%%%%
\begin{figure}[hbtp]
\begin{center}
\includegraphics[width=14cm,height=6cm]{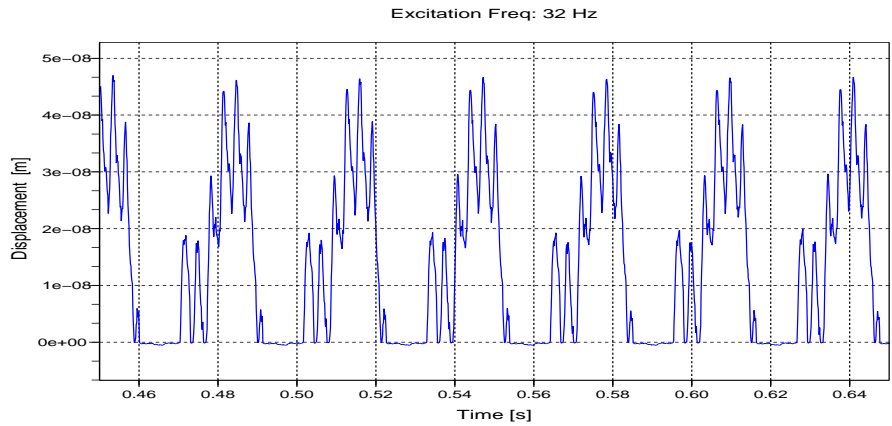}
\caption{The predicted displacements for an excitation at $32$ Hz; the displacement
is measured immediately above the support. The high unilateral stiffness yields almost a positive displacement.}\label{d_32}
\end{center}
\end{figure}
\begin{figure}[hbtp]
\begin{center}
\includegraphics[width=14cm,height=6cm]{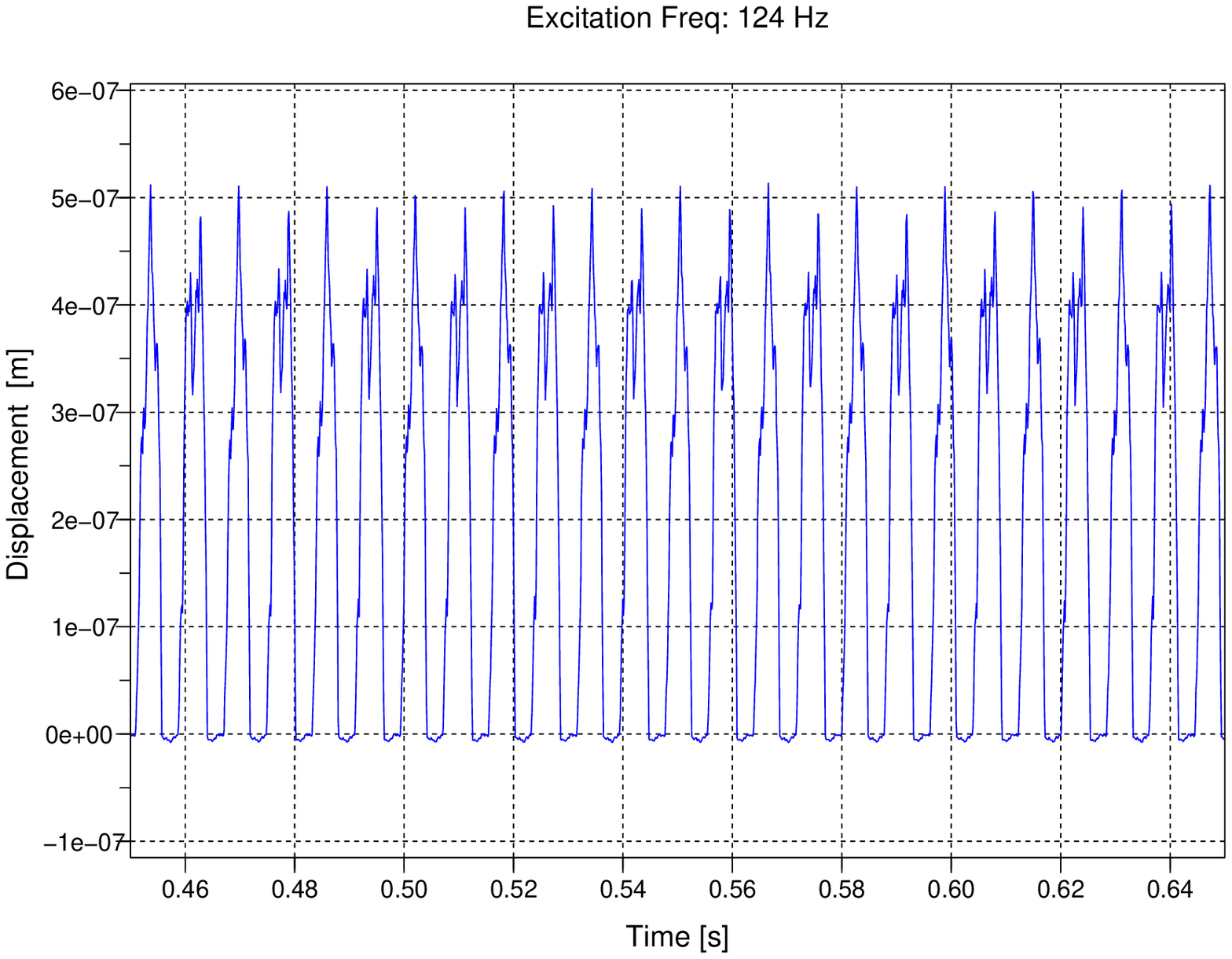}
\caption{The predicted displacements for an excitation at $124$ Hz, the displacement
is measured immediately above the support.}\label{d_124}
\end{center}
\end{figure}
\begin{figure}[hbtp]
\begin{center}
\includegraphics[width=14cm,height=6cm]{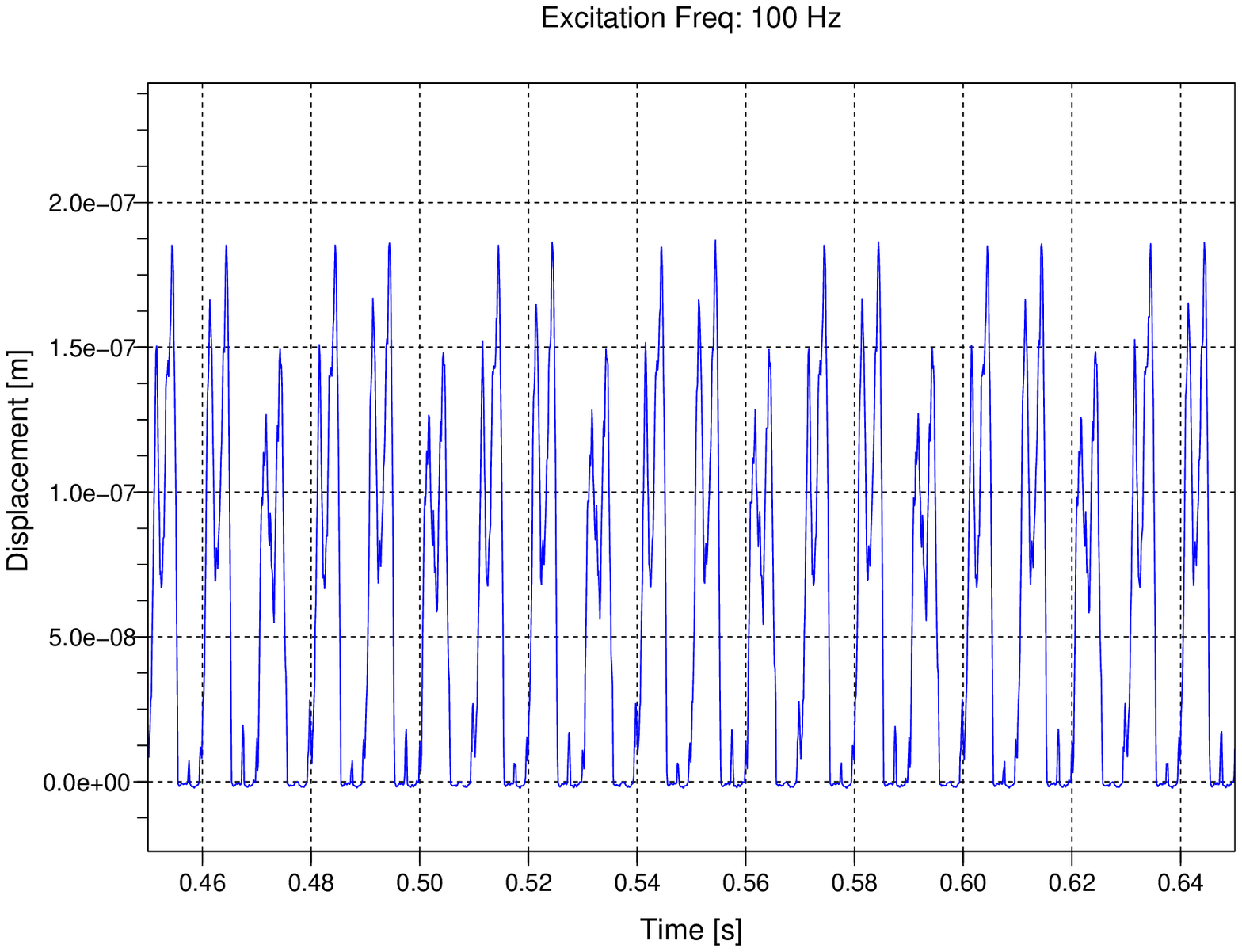}
\caption{The predicted displacements for an excitation at $100$ Hz, the displacement
is measured immediately above the support.}\label{d_100}
\end{center}
\end{figure}
%%%%%%%%%%%%%%%%%%%%%%%%%%%%%%%%%%%%%%%%%%%%%%%%%%%%%%%%%%%%%%%%%%%%%%%%%%
\begin{figure}[hbtp]
\begin{center}
\includegraphics[width=14cm,height=7cm]{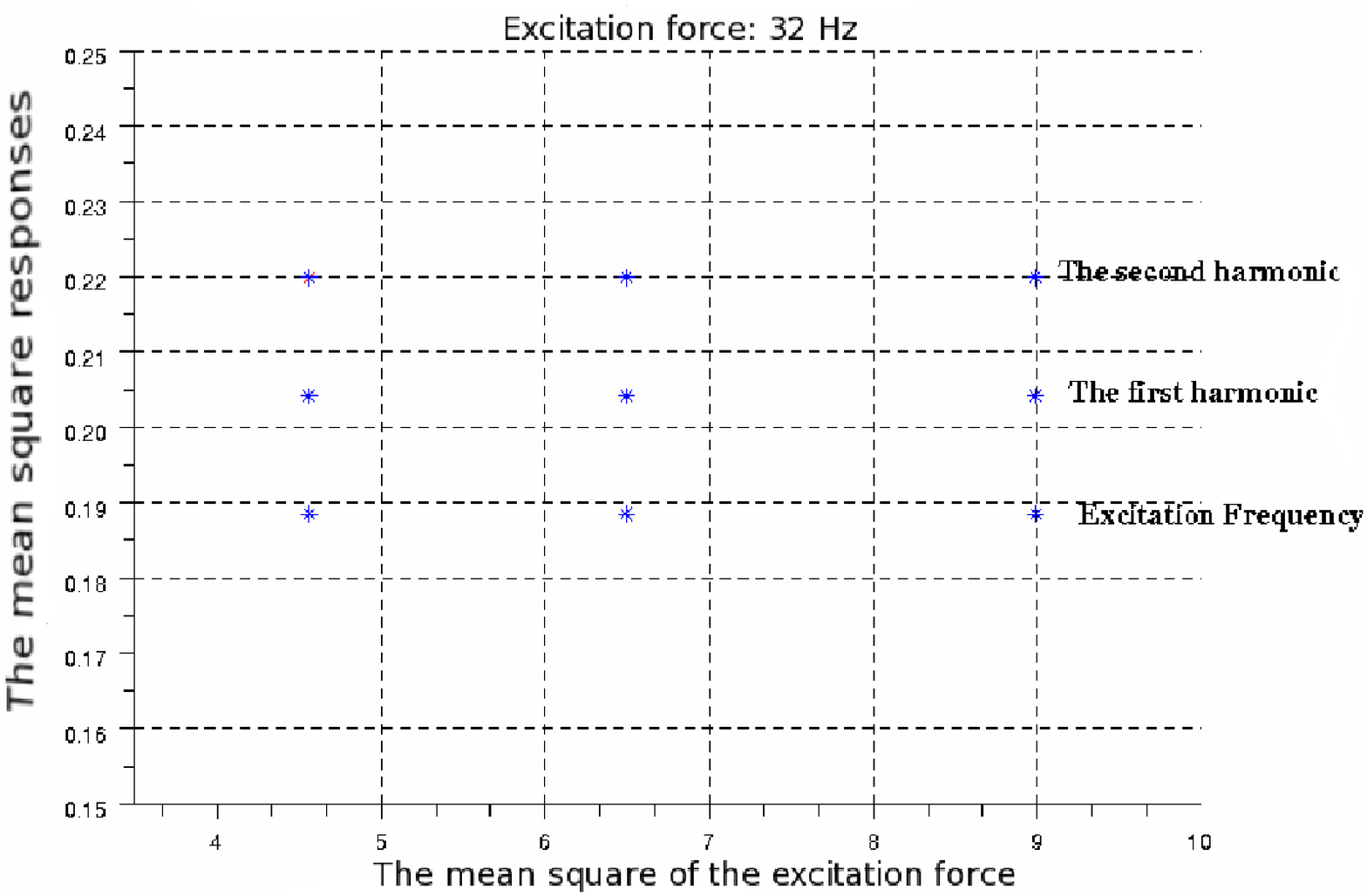}
\caption{The normalized mean square responses (mean square displacement divided by the mean square excitation force) in each harmonic for inputs at three different mean square force levels. The excitation frequency is $32$ Hz, the acceleration is measured immediately above the support.}
\label{ed}
\end{center}
\end{figure}
%%%%%%%%%%%%%%%%%%%%%%%%%%%%%%%%%%%%%%%%%%%%%%%%%%%%%%%%%%%%%%%%%%%%
\begin{figure}[hbtp]
\begin{center}
\includegraphics[width=14cm,height=6cm]{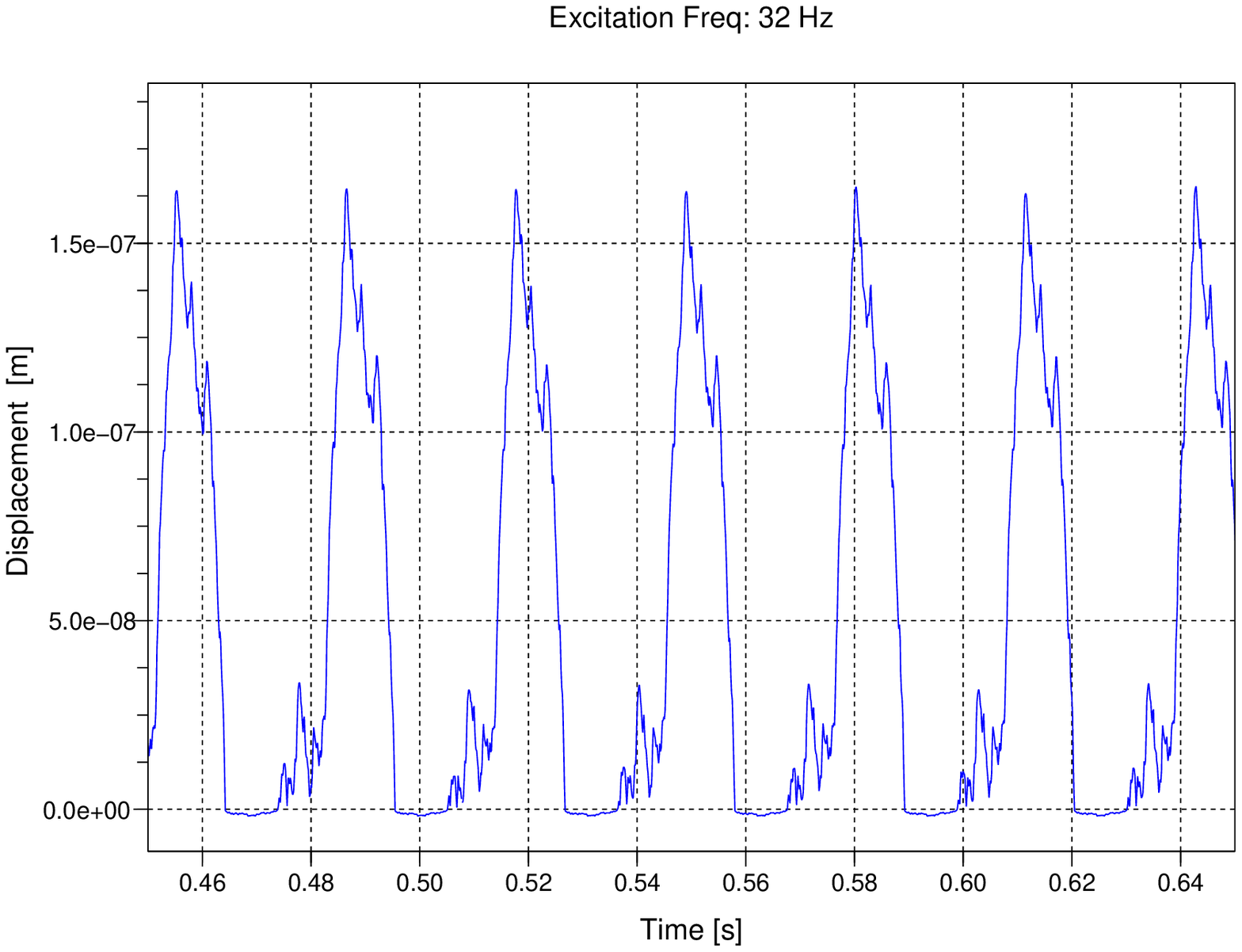}
\caption{The predicted displacements for a sine excitation at $32$ Hz, The acceleration  magnitude is $a=1$ $m/s^2$. The displacement
is measured immediately above the support.}
\label{d_sine_32}
\end{center}
\end{figure}

\begin{figure}[hbtp]
\begin{center}
\includegraphics[width=14cm,height=6cm]{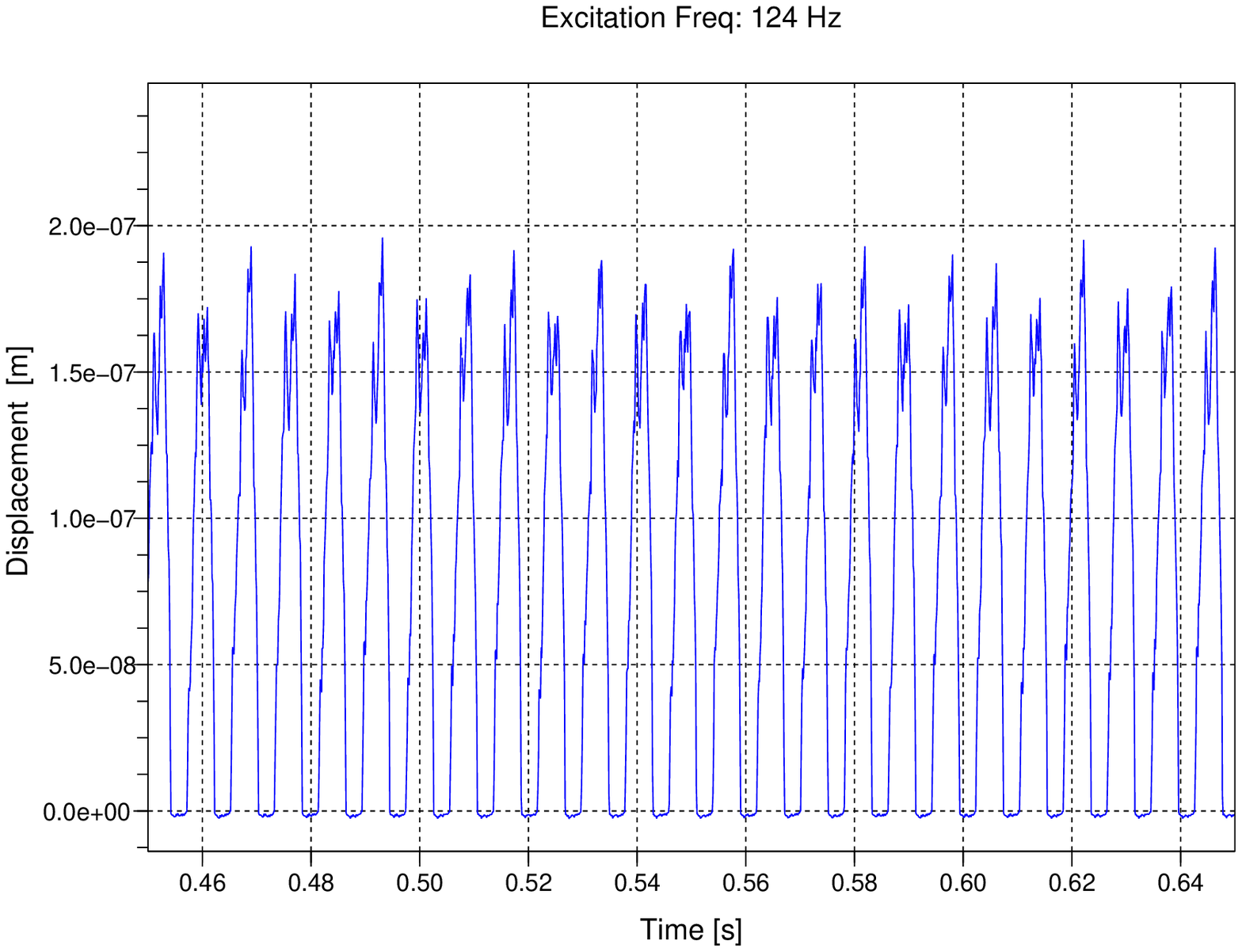}
\caption{The predicted displacements for a sine excitation at $124$ Hz. The acceleration  magnitude is $a=1$ $m/s^2$. The displacement
is measured immediately above the support.}
\label{d_sine_124}
\end{center}
\end{figure}

\begin{figure}[hbtp]
\begin{center}
\includegraphics[width=14cm,height=6cm]{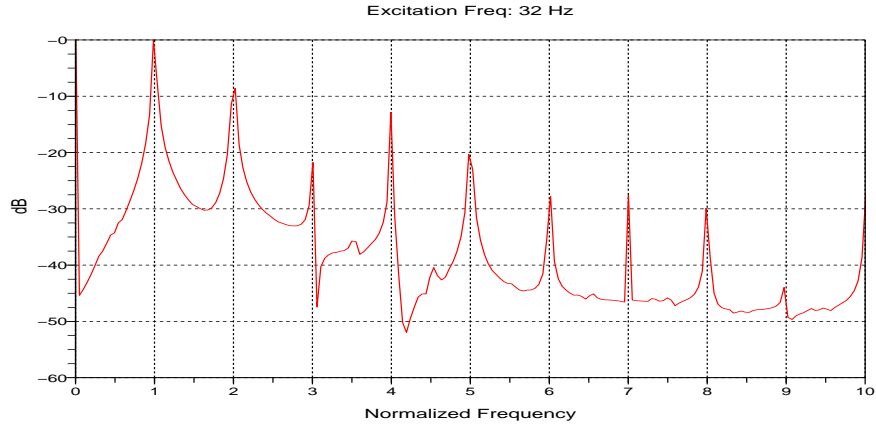}
\caption{The frequency content of the predicted numerical displacement for sine  excitation at $32$ Hz, the displacement is measured immediately above the support. The excitation frequency is split into all its harmonics.}
\label{fft_sine_32}
\end{center}
\end{figure}

%%%%%%%%%%%%%%%%%%%%%%%%%%%%%%%%%%%%%%%%%%%%%%%%%%%%%%%%%%%%%%%%
\begin{figure}[hbtp]
\begin{center}
\includegraphics[width=14cm,height=6cm]{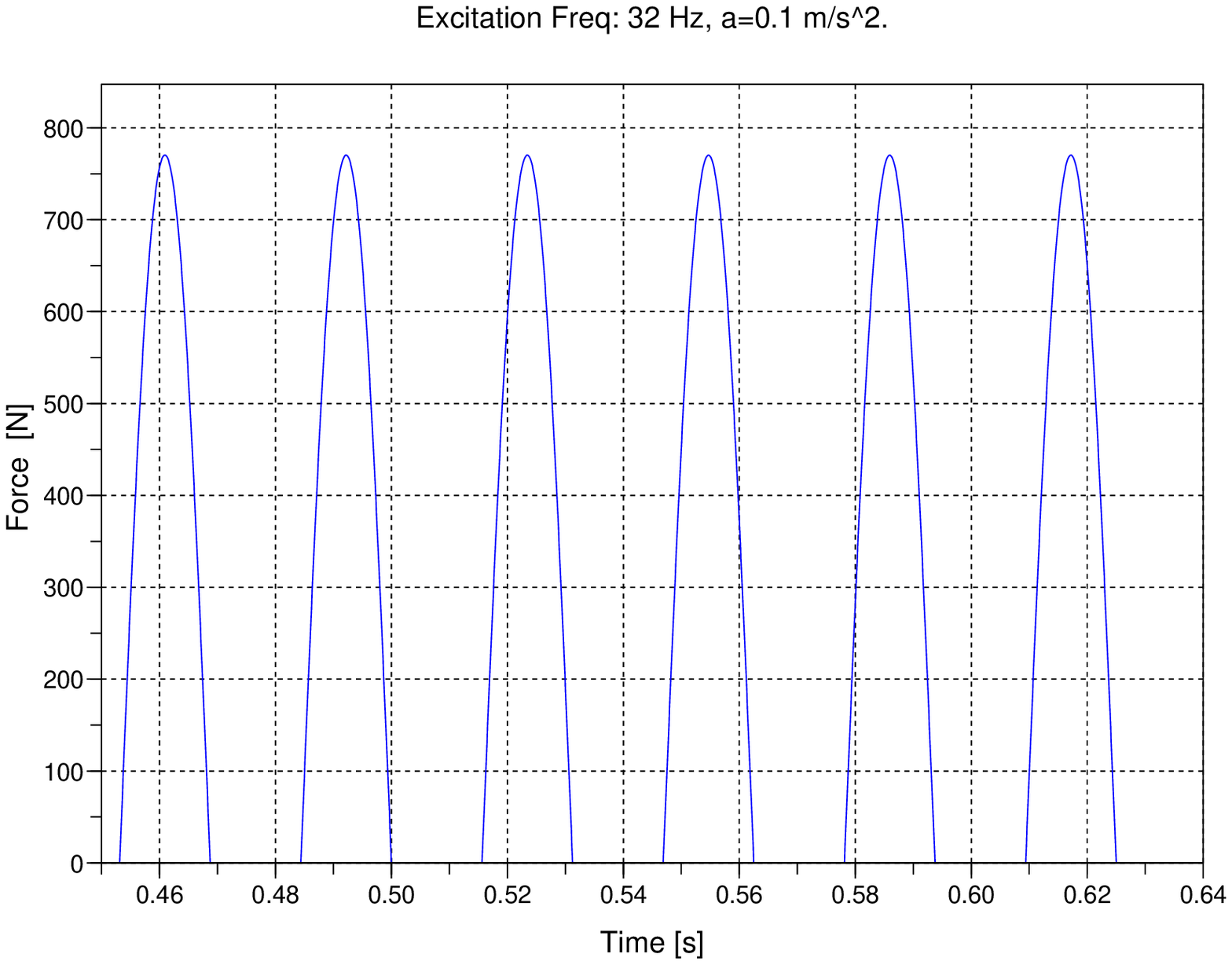}
\caption{The predicted elastic force of the spring support for an excitation at $32$ Hz. The acceleration  magnitude is $a=0.1 m/s^2$ and the spring is only in contact at times where the beam displacement is negative.}
\label{force_sine_32}
\end{center}
\end{figure}

\begin{figure}[hbtp]
\begin{center}
\includegraphics[width=14cm,height=6cm]{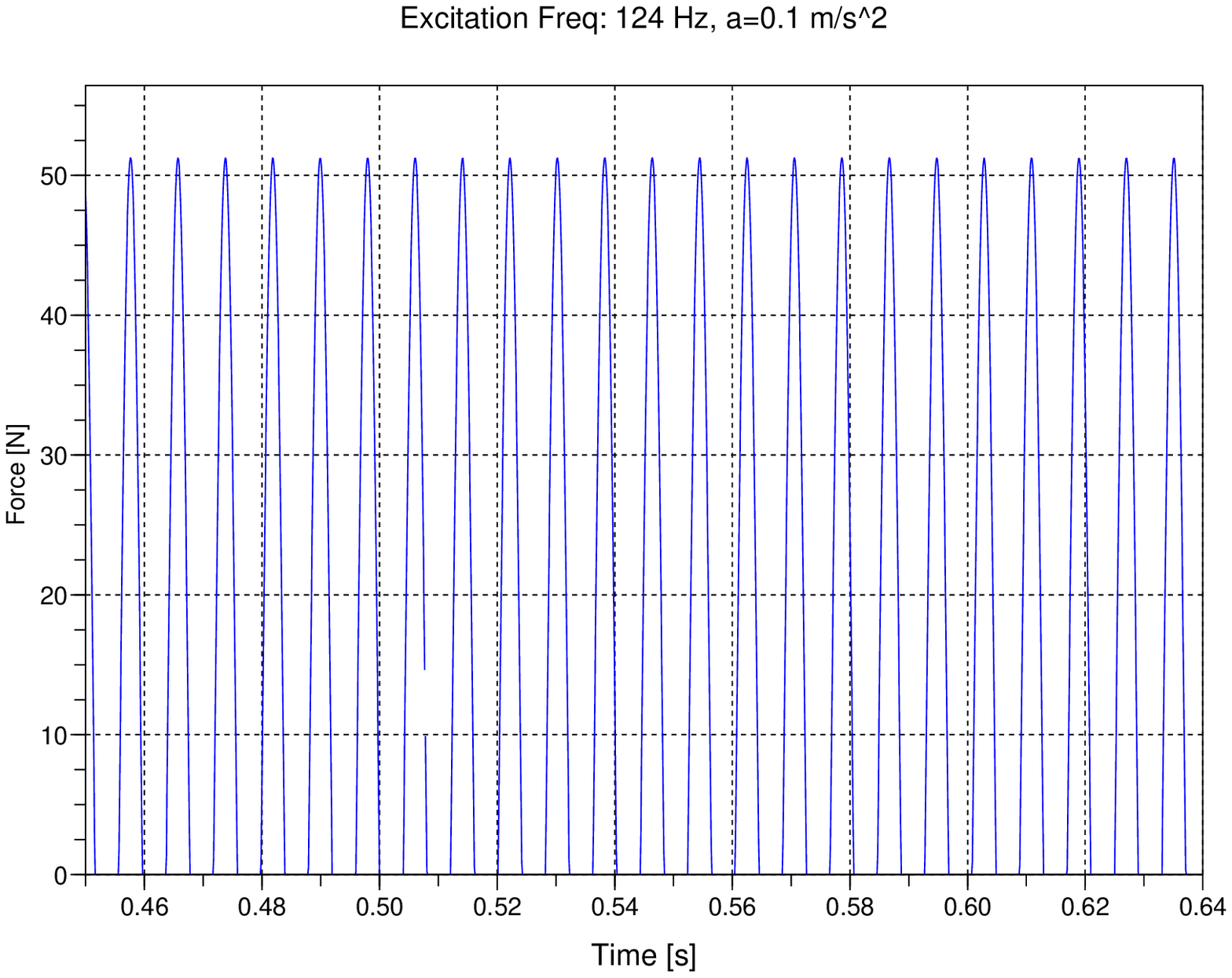}
\caption{The predicted elastic force of the spring support for an excitation at $124$ Hz. The acceleration  magnitude is $a=0.1 m/s^2$ and the spring is only in contact at times where the beam displacement is negative.}
\label{force_sine_124}
\end{center}
\end{figure}
%%%%%%%%%%%%%%%%%%%%%%%%%%%%%%%%%%%%%%%%%%%%%%%%%%%%%%%%%%%%%%%%%%%%%%%%%%%%

\begin{figure}[hbtp]
\begin{center}
\includegraphics[width=14cm,height=6cm]{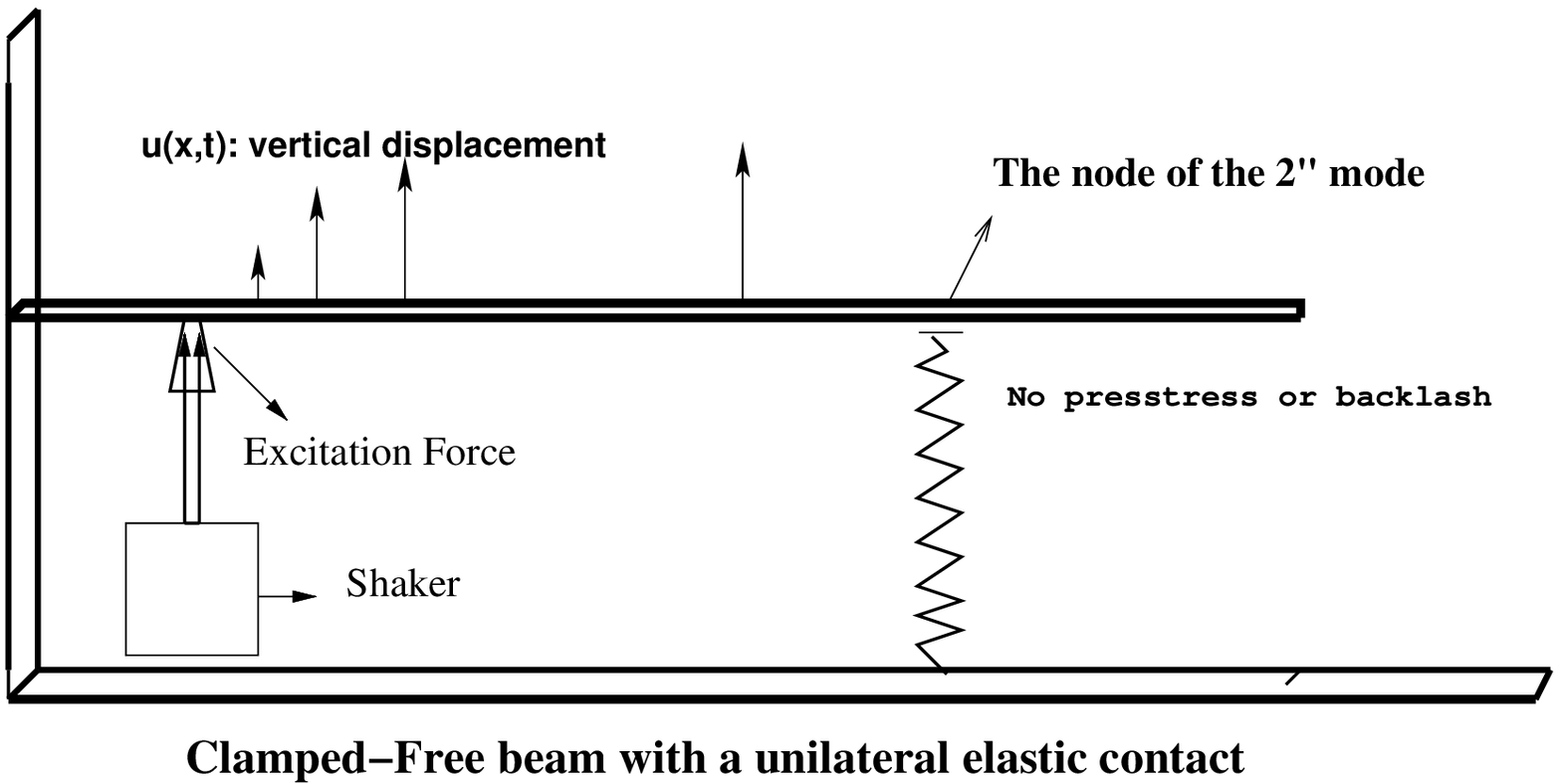}
\end{center}
\caption{\small{beam system with an unilateral spring under a periodic excitation}} \label{2}
\end{figure}
\newpage
\begin{figure}[hbtp]
\begin{center}
\includegraphics[width=14cm,height=5cm]{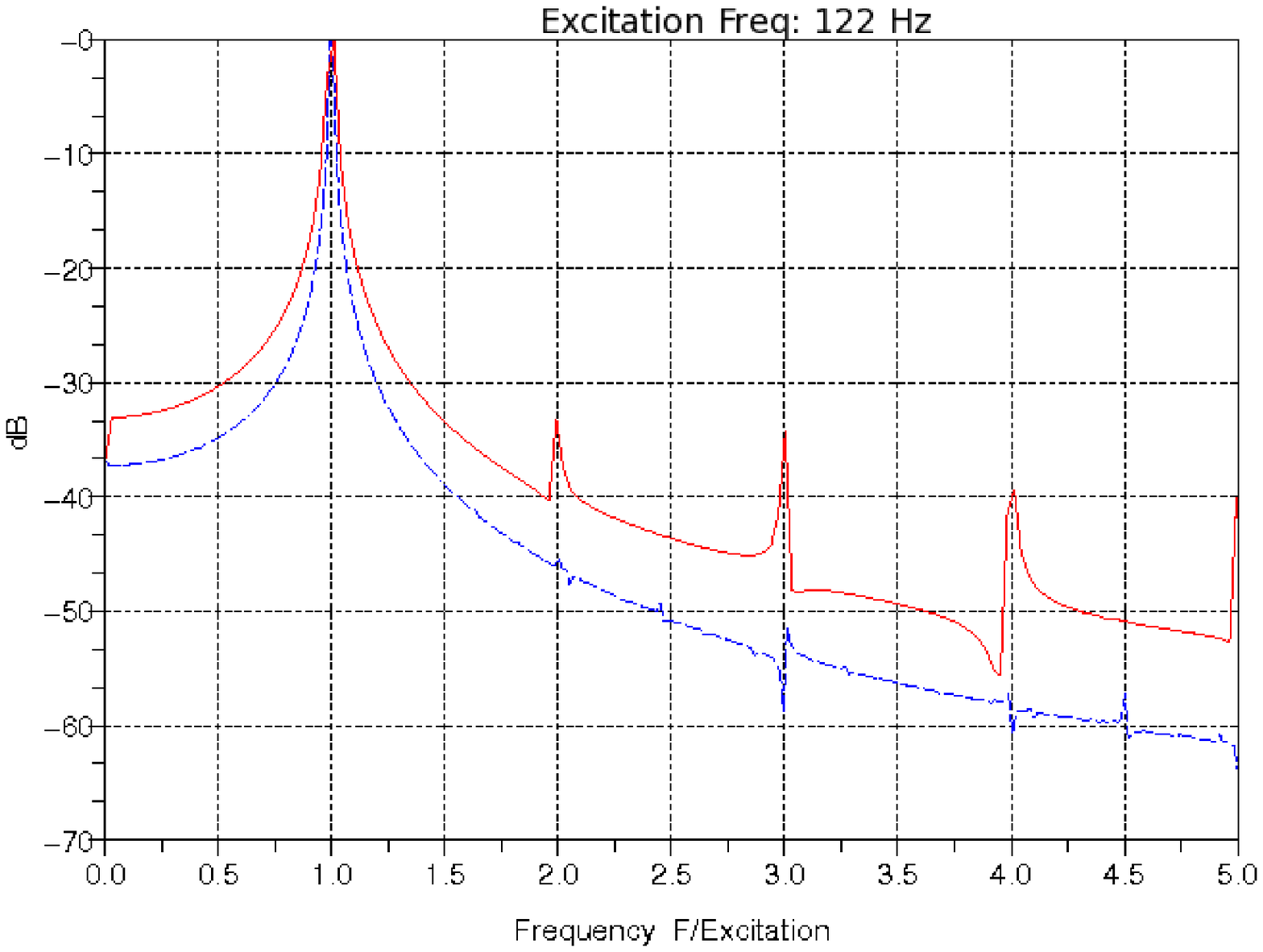}
\caption{Predicted (solid) and measured displacements (dB) for an excitation at $122$ Hz applied to the beam with unilateral support stiffness. The displacement
is measured immediately above the support and the frequency axis is normalized by the excitation frequency.}
\label{fft_new_122}
\end{center}
\end{figure}
%%%%%%%%%%%%%%%%%%%%%%%%%%%%%%%%%%%%%%%%%%%%%%%%%%%%%%%%%%%%%%%%%%%%%%%%%%%%%%%%%%%%%%%%%%%%%%%%%%%%%%%%%%%\newpage%%%%%%%%%
\newpage
\begin{figure}[hbtp]
\begin{center}
\includegraphics[width=14cm,height=7cm]{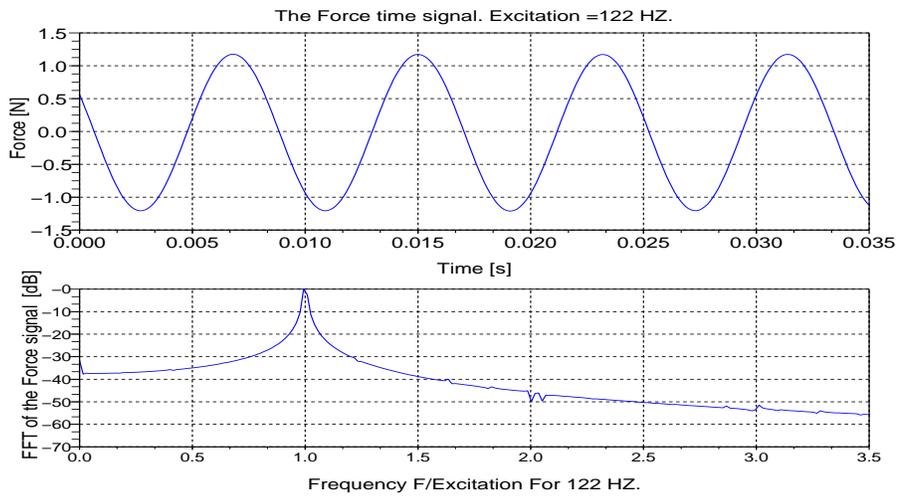}
\caption{The time signal and its FFT of the input force for an excitation at $122$ Hz.}
\label{force_new}
\end{center}
\end{figure}
%\n\newpageewpage
\newpage
\begin{figure}[hbtp]
\begin{center}
\includegraphics[width=14cm,height=7cm]{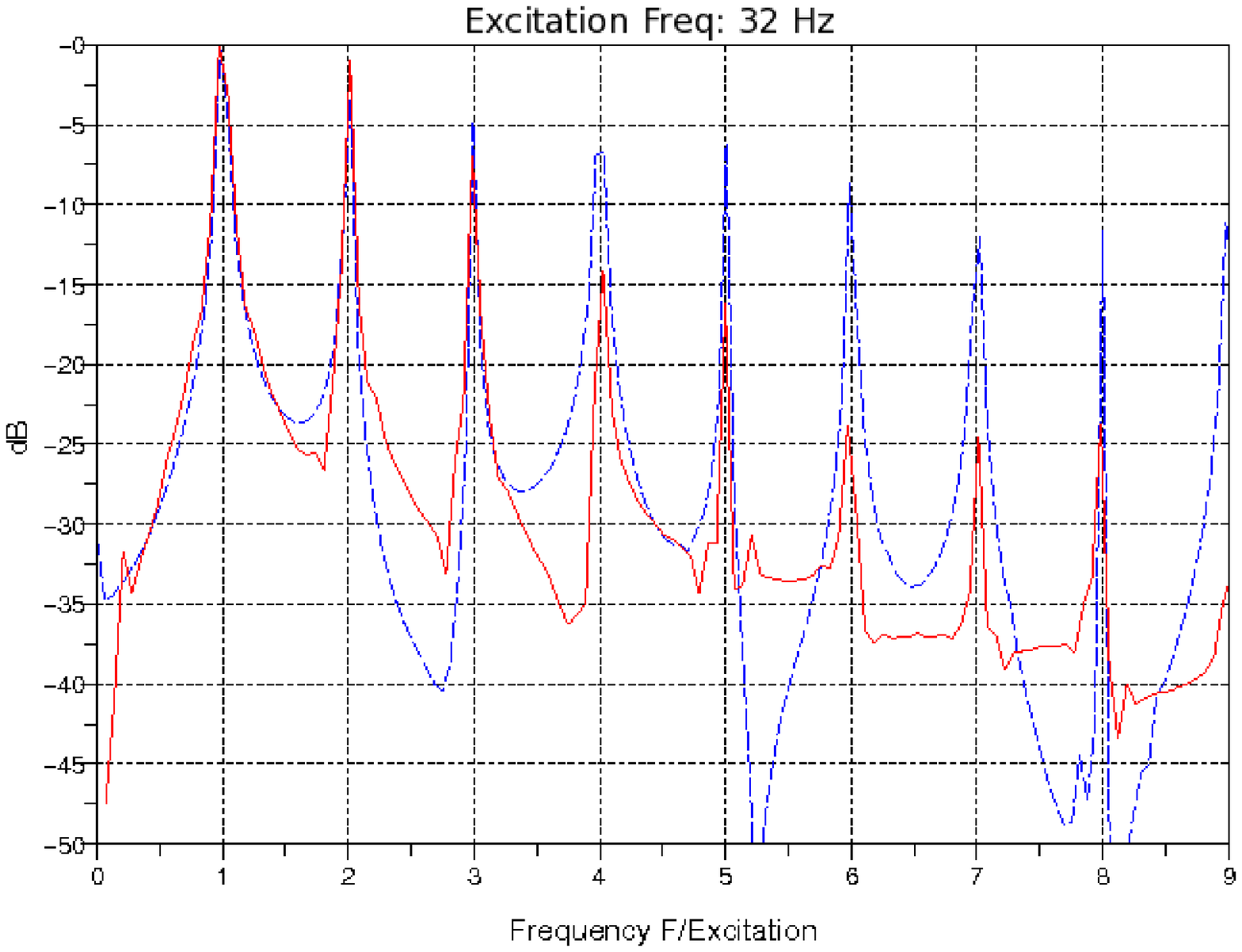}
\caption{Predicted (solid) and measured displacements (dB) for an excitation at $32$ Hz applied to the beam with unilateral support stiffness. The displacement
is measured immediately above the support and the frequency axis is normalized by the excitation frequency.}
\label{fft_new_32}
\end{center}
\end{figure}
%%%%%%%%%%%

\end{document}